\PassOptionsToPackage{pdfpagelabels=false}{hyperref} 
\documentclass[fleqn,usenatbib]{mnras}

\usepackage{newtxtext,newtxmath}

\usepackage[T1]{fontenc}
\usepackage{ae,aecompl}

\makeatletter
\newlength{\abovecaptionskip}%
\makeatother

\usepackage{amsmath,amstext}
\usepackage{graphicx}	
\usepackage[figure,figure*]{hypcap}
\usepackage{enumerate}
\usepackage{xspace} 
\usepackage{multirow}
\usepackage[flushleft]{threeparttable}
\usepackage{booktabs}
\usepackage{adjustbox}

\usepackage{marginnote}

\usepackage{soul}
\setstcolor{red}

\makeatletter
\newcommand{\HI}{{\text{H\MakeUppercase{\romannumeral 1}}}\xspace}
\newcommand{\HII}{{\text{H\MakeUppercase{\romannumeral 2}}}\xspace}
\newcommand{\HeI}{{\text{He\MakeUppercase{\romannumeral 1}}}\xspace}
\newcommand{\HeII}{{\text{He\MakeUppercase{\romannumeral 2}}}\xspace}
\newcommand{\HeIII}{{\text{He\MakeUppercase{\romannumeral 3}}}\xspace}
\newcommand{\Lya}{\ifmmode{\mathrm{Ly}\alpha}\else Ly$\alpha$\xspace\fi}
\newcommand{\Htwo}{\ifmmode{\mathrm{H}_2}\else H$_2$\xspace\fi}
\newcommand{\MBII}{{\text{MB\MakeUppercase{\small\expandafter{\romannumeral 2}}}}\xspace}

\newcommand{\Rmnum}[1]{\expandafter\@slowromancap\romannumeral #1@}

\newcommand{\h}[1]{\ensuremath{h^{-1}} #1 }
\newcommand{\diff}{\ensuremath{\; \text{d}}}

\newcommand{\abb}[2]{{#1}{\small \uppercase\expandafter{\romannumeral #2}}}
\newcommand{\abbm}[2]{\ensuremath{{\rm #1}{\small\rm\scriptstyle \uppercase\expandafter{\romannumeral #2}}}}

\title[The Epoch of IGM heating by early sources of X-rays]{The Epoch of IGM heating by early sources of X-rays}

\author[M.B.~Eide et al.]{Marius B. Eide$^{1}$\thanks{Fulbright Fellow. E-mail: eide@MPA-Garching.MPG.DE}
Luca Graziani$^{1,2,3}$,
Benedetta Ciardi$^{1}$, 
Yu Feng$^{4}$,
\newauthor
Koki Kakiichi$^{1,5}$
and Tiziana Di Matteo$^{6}$
\\
$^{1}$Max-Planck-Institut f\"ur Astrophysik, Karl-Schwarzschild-Stra\ss e 1, 85741 Garching, Germany\\
$^{2}$INAF / Osservatorio Astronomico di Roma, Via di Frascati 33, 00078 Monte Porzio Catone, Italy\\
$^{3}$Scuola Normale Superiore, Piazza dei Cavalieri 7, 56126, Pisa, Italy\\
$^{4}$Berkeley Center for Cosmological Physics Campbell Hall 341, University of California, Berkeley CA 94720, United States\\
$^{5}$Department of Physics and Astronomy, University College London, London, WC1E 6BT, UK\\
$^{6}$McWilliams Center for Cosmology, Physics Department, Carnegie Mellon University, Pittsburgh, PA 15213, USA
}

\date{Accepted XXX. Received YYY; in original form ZZZ}

\pubyear{2017}

\begin{document}
\label{firstpage}
\pagerange{\pageref{firstpage}--\pageref{lastpage}}
\maketitle

\begin{abstract}
    Observations of the 21~cm line from neutral hydrogen indicate that an Epoch of Heating (EoH) might have preceded the later Epoch of Reionization  (EoR).
    Here we study the effects on the ionization state and the thermal history of the Intergalactic Medium (IGM) during the EoH induced by different assumptions on  ionizing sources in the high redshift Universe:  (\textit{i}) stars, (\textit{ii}) X-ray binaries (XRBs), (\textit{iii}) thermal bremsstrahlung of the hot Interstellar Medium (ISM), and (\textit{iv}) accreting nuclear black holes (BHs).
    To this aim, we post-process outputs from the (100 $h^{-1}$ cMpc)$^3$ hydrodynamical simulation MassiveBlack-II with the cosmological 3D radiative transfer code \texttt{CRASH}, which follows the propagation of UV and X-ray photons, computing the thermal and ionization state of hydrogen and helium through the EoH.
    We find that stars determine the fully ionized morphology of the IGM, while the spectrally hard XRBs pave way for efficient subsequent heating and ionization by the spectrally softer ISM.
    With the seeding prescription in MassiveBlack-II, BHs do not contribute significantly to either ionization or heating.
    With only stars, most of the IGM remains in a cold state (with a median $T=11$~K at $z=10$), however, the presence of more energetic sources raises the temperature of regions around the brightest and more clustered sources above that of the CMB, opening the possibility to observing the 21~cm signal in emission.  
\end{abstract}

\begin{keywords}
cosmology:\ dark ages, reionization, first stars -- radiative transfer
\end{keywords}

\section{Introduction}
A central, outstanding question in cosmology is the evolution and nature of the hydrogen and helium constituting the intergalactic medium (IGM), particularly at high redshift.
Hydrogen was expected to reveal itself either as a deep trough blueward of the \HI 21~cm line \citep{Davies1964} or the \HI \Lya line \citep{Gunn1965} in the spectra of distant, bright radio galaxies \citep{Field1962,Penzias1968,Allen1969}.
However, this did not happen, leading to the conclusion that the IGM must presently be highly ionized and hot.
Not until recently was the \HI trough finally detected, but at higher redshifts \citep[e.g.][]{Fan2000,Fan2006,McGreer2015}.
The IGM must once have been neutral, and then have undergone a phase transformation, the \textit{Cosmic Dawn}, into its present-day state.
Persisting questions are \textit{how} this transformation occurred, and \textit{what} drove it.
Answering them will shed light on the physical properties of the majority of the baryonic Universe, and open doors to understanding the interplay between the IGM and the large- and small-scale structures that stem from it.
The IGM is thus a highly uncharted terrain (see e.g.~\citealt{Meiksin2009} for a review).

In this work, we investigate the properties and signatures of the sources of ionizing, and possibly heating, radiation
that flared up as structures formed, marking the transition from the \textit{Dark Ages} to the Cosmic Dawn \citep{Rees1996}.
Early cosmological 21 cm surveys have begun to report that heating may precede full ionization, resulting in a separate \textit{Epoch of Heating} (EoH) preceding the \textit{Epoch of Reionization} (EoR).
The recent constraint on the global signal with SARAS 2 \citep{Singh2017} disfavors
late heating, disentangling the EoH from the EoR \citep{Fialkov2014}.
These observations also align themselves with other surveys, as the early results of \citet{Bowman2008}, that were followed by upper limits on the scale and magnitude of brightness temperature fluctuations (e.g.~from GMRT, \citealt{Paciga2011}; MWA, \citealt{Dillon2014,Ewall-Wice2016}; PAPER, \citealt{Parsons2014,Ali2015}; and LOFAR, \citealt{Patil2017}).
However, these surveys are impeded at very high redshift by the ionospheric contamination from Earth, which is nearly fully opaque to redshifted 21~cm signal.
To mitigate for this, a seminal Chinese-Dutch mission scheduled for 2018 plans to obtain the global 21~cm signal from the far side of the Moon\footnote{https://www.isispace.nl/dutch-radio-antenna-depart-moon-chinese-mission/, accessed 13 July 2017.}.

Besides 21~cm tomography, which is still in its infancy after its revival as a probe by \cite{Madau1997}, there are other observational means for examining the ionization and thermal state of the IGM.
The spectra of bright, single sources \citep[as quasars or gamma ray bursts, see e.g.~review by][]{Ciardi2005} reveal the properties of the intervening IGM along the line-of-sight.
The increasing abundance of neutral \HI at earlier $z$ has been indicated
from large-scale surveys, as well (e.g.~\citealt{Matthee2015,Zheng2017,Ouchi2017} or see review by \citealt{Dijkstra2014}).
Hydrogen ionization is accompanied by production of free electrons.
Satellite-based experiments \citep{Komatsu2011,PlanckCollaboration2016} have provided tight constraints on the scattering of the cosmic microwave background (CMB) off these electrons, as the CMB becomes dampened and linearly polarized.
These observations are however in slight tension, as the continuing lowering of the Thomson scattering optical depth requires a later completion of cosmic reionization.

We do not expect a single source type of heating and ionization to be able to account for the diversity in the observed signatures.
\cite{Kakiichi2017} instead found the interplay between different ionizing sources to be essential in establishing the thermal and ionization state of the IGM and to leave different signatures.
Possible viable candidates include:

(\textit{i}) Stars residing within galaxies.
These are thought to be the primary drivers of reionization, as argued by \cite{Madau1999}.
The recent low Thomson scattering optical depth reported by the \cite{PlanckCollaboration2016}, combined with constraints from Hubble \citep{Robertson2015}, has rejuvenated the interest in their effect on reionization.
Structure formation is also affected by the thermal and ionizing feedback from stars, as discussed by e.g.~\cite{Couchman1986}, \cite{Cen1992}, and \cite{Fukugita1994}.
Depending on their initial mass function, mass and metallicity, they will produce copious amounts of ionizing photons, as shown in the review by \cite{Ciardi2005}. 

(\textit{ii}) Accreting nuclear black holes (or simply black holes, BHs). Hereafter this  term will refer to active galactic nuclei (AGN) and the brighter QSOs.
While these were early candidates for IGM ionization \citep{Arons1969,Rees1969,Bergeron1970},
assessing their importance is strongly hampered by lacking data on the evolution of the faint end of the QSO luminosity function (QLF) at higher redshifts. 
With the detection of 22 faint AGNs by \cite{Giallongo2015}, the possibility of them producing amounts of ionizing photons sufficient to drive reionization independently from stars, was once again examined.
\cite{Madau2015} and \cite{Mitra2016} found this to be possible, but it would also doubly ionize helium before redshift 4, in contrast to recent observations of an extended \HeII reionization at $z\sim 2.8$ \citep{Worseck2016}. 
\cite{Qin2017} report that the models of \cite{Madau2015} and \cite{Mitra2016} overpredict the contribution from AGN, and do not find them to dominate over stars, a conclusion also shared by \cite{Hassan2017} and \cite{Onorbe2017}.
Updated QLF by \cite{Parsa2017} or \cite{Onoue2017} also indicate the contribution of BHs to be insufficient to drive reionization.
However, the contribution from AGNs were found by \cite{Chardin2017} and \cite{DAloisio2017} to be vital to the large scale opacity fluctuations observed in the \Lya forest, which was strengthened by the recent observations of a through spanning $240 h^{-1}$ cMpc by \cite{Barnett2017}.

(\textit{iii}) Galactic X-ray binary systems  comprising a neutron star or a black hole devouring a companion star. Among such systems,  the majority of the ionizing luminosity at high-$z$ originates from massive (HMXBs) rather than low-mass (LMXBs) binary systems \citep{Mirabel2011,Fragos2013a,Fragos2013,Madau2017,Sazonov2017}. \cite{Mineo2012,Mineo2012a} found the spectra of XRBs to be too hard to account for the soft X-ray flux of galaxies, while they become dominant at higher energies.

(\textit{iv}) Thermal bremsstrahlung in the diffuse gas of the heated interstellar medium (ISM). 
The heating mechanism of the diffuse gas has been a  topic under investigation for decades (see e.g.~the review by~\citealt{Fabbiano1989}). 
Interactions between supernova driven galactic superwinds and clouds \citep[e.g.~][]{Chevalier1985} is the preferred explanation \citep{Pacucci2014}. 
Shocked gas in the galactic halo and disk  \citep[e.g.~][]{Suchkov1994}, as well as hot galactic winds \citep[e.g.~][]{Strickland2000} are also processes that could yield predominantly soft X-rays.

(\textit{v}) Other candidates, which  will not be examined further in this paper because of their secondary role, are e.g.~ low energy cosmic rays \citep{Ginzburg1966,Nath1993, Sazonov2015,Leite2017}, self-annihilation or decay of dark matter (e.g. \citealt{Liu2016}) and plasma beam instabilities in TeV blazars (e.g. \citealt{Chang2012,Puchwein2012}).

There are different approaches for simulating the physics of the Cosmic Dawn.
Ideally, structure formation can be coupled to radiative transfer
(e.g. \citealt{Gnedin2014,So2014,OShea2015,Ocvirk2016,Pawlik2017,Semelin2017}).
However, this is computationally expensive and requires simplifications. For example helium physics is often not treated, or the number of frequency bins used may be limited to a few, albeit the cross section of hydrogen varies approximately as $\nu^{-3}$.
The spatial extent is also often well below the required $\sim 100 \h$ cMpc needed to provide a consistent reionization history \citep{Iliev2014}.
Semi-numerical codes as 21CMFAST \citep{Mesinger2011} or SimFast21 \citep{Santos2010,Hassan2016} can be applied on large scales using an excursion-set formalism \citep{Furlanetto2004}.
This allows for fast parameter-space exploration, but may lack the treatment of some important physical processes, such as helium reionization, temperature
evolution, partial ionization and X-ray implementation. 
Nevertheless, this approach has provided to be rewarding in studies of possible consequences of radiative feedback on structure formation and its 21~cm signatures \citep[see e.g.~][]{Fialkov2013}.
Radiative post-processing of hydrodynamical or N-body simulations is also possible \citep[e.g.~][]{Baek2010,Ciardi2012,Graziani2015,Ross2017} and allows for the dedication of computational power to e.g.~treat heating, multifrequency radiative transfer (RT), as well as the effects of helium.

Here we examine the effects of \textit{stars, BHs, XRBs} and the \textit{ISM} on both the thermal and ionization states of the IGM during the EoH.
Our approach is the following: we rely on the hydrodynamical cosmological structure formation simulation MassiveBlack-II  \citep[\MBII, described in \S\ref{ssec:cosmo}]{Khandai2015}, which includes baryonic physics and feedback processes, to provide us with the physical environment of the IGM (temperature and gas density), as well as
the location and properties of the sources.
We assign spectra and ionizing luminosity to star forming particles based on their star-formation rates, masses, ages and metallicities, and to black hole particles based on their accretion rates.
We then perform radiative post-processing of the \MBII simulations with the  cosmological radiative transfer code \texttt{CRASH}
(see \S\ref{ssec:rad_trans}) which gives us the evolution of the thermal and ionization state of the IGM.
The properties of the ionizing sources are based on empirical relations introduced in \S\ref{ssec:source_modelling}.
We present our findings in \S\ref{sec:results} and relate them to comparable studies also summarizing our conclusions in \S\ref{sec:discussion} and \S\ref{sec:conclusion}.

\section{Methodology}
\label{sec:methods}
Here, we describe how we combine the outputs of the cosmological hydrodynamical simulation \MBII (\S\ref{ssec:cosmo}) with population synthesis modeling of ionizing sources (\S\ref{ssec:source_modelling}), and finally perform multifrequency radiative transfer in the same cosmological volume with \texttt{CRASH} (\S\ref{ssec:rad_trans}).

\subsection{Cosmological hydrodynamic simulation}
\label{ssec:cosmo}

\abb{MB}{2} (\citealt{Khandai2015}, K15) is a high resolution cosmological SPH simulation tracking stellar populations, galaxies, accreting and dormant black holes, as well as their properties (as position, age, metallicity, mass, accretion rate, star formation rate). The simulation has been run using \texttt{P-GADGET}, a newer version of \texttt{GADGET-3} \citep[see][for an earlier version]{Springel2005b}.
It accounts for baryonic physics and feedback effects of the sources on their environment following earlier works in its approach to feedback \citep{DiMatteo2008,Croft2009,Degraf2010,DiMatteo2012}, sub-grid treatment of star formation \citep{Springel2003}, and seeding and evolution of black holes \citep{DiMatteo2005,Springel2005a}.
The simulation has a box length of $100h^{-1}$ cMpc and is
performed in the WMAP7 $\Lambda$CDM cosmology \citep{Komatsu2011}\footnote{$\sigma_8=0.816$, $n_s=0.968$, $\Omega_\Lambda=0.725$, $\Omega_{\rm m}=0.275$, $\Omega_{\rm b}=0.046$, $h=0.701$}, using $2\times1792^3$ particles of mass $m_{\rm DM}=1.1\times10^7 h^{-1} {\rm M}_\odot$ and $m_{\rm gas}=2.2\times10^6 h^{-1} {\rm M}_\odot$ for dark matter and gas, respectively. The adopted gravitational softening length is $1.85 h^{-1}$ ckpc.
As a reference, we have a total of 28 and 169,520 
haloes hosting at least a star particle at $z=18$ and $10$, respectively, where the lowest dark matter halo masses are $2 \times 10^8 h^{-1} {\rm M}_\odot$ and $9 \times 10^7 h^{-1} {\rm M}_\odot$.
The highest dark matter halo masses are $2 \times 10^9 \h {\rm M}_\odot$ and $1 \times 10^{11} \h {\rm M}_\odot$ at $z=18$ and $10$, respectively.
The BHs form from seeds of $5 \times 10^5 h^{-1} {\rm M}_\odot$, growing with accretion rates in the range $(10^6$--$10^7) h^{-1} {\rm M}_\odot {\rm Gyr}^{-1}$ to a maximum mass of $1.4 \times 10^{6} h^{-1} {\rm M}_\odot$
at $z=10$. 
The first seed black hole is found at $z=13$, while
we have seventeen BHs at $z=10$.
We refer the reader to K15 for more details on the simulation.

We employ six snapshots from \abb{MB}{2}, covering the evolution between redshifts $z=18$ and $z=10$,
each describing the instantaneous state of the simulation.  
The particle distribution of each snapshot is mapped  onto a Cartesian grid of $N_c^3$ cells to create maps  of gas number density and temperature, as well as location and properties of the ionizing sources (see \S\ref{ssec:source_modelling} for a detailed description of how we convert \MBII data to ionizing sources). 
The reference value is $N_c=256$, corresponding to a spatial resolution of $391 h^{-1}$~ckpc.
After gridding, the 45 (292,685)
star particles present at $z=18$ (10) are reduced to 26
(56,702). These are effectively our sources. When more than one particle ends up in a cell, their properties  are summed up (see following section).
The gas density is converted to hydrogen and helium number densities by assuming a number fraction $X = 0.92$ and $Y = 0.08$, respectively, and no metals.

\subsection{Cosmological radiative transfer simulation}
\label{ssec:rad_trans}
The radiative transfer of ionizing photons is
performed by post-processing the outputs of \abb{MB}{2} with the multifrequency Monte Carlo ray-tracing code \texttt{CRASH} \citep{Ciardi2001, Maselli2003, Maselli2009, Graziani2013}, which calculates the ionization state of hydrogen and helium, as well as the gas temperature in each grid cell traversed by photons emitted by a radiation source.
The version of \texttt{CRASH} employed here features a self-consistent treatment of UV and soft X-ray photons, in which X-ray ionization and heating as well as detailed secondary electron physics are included (Graziani et al. in prep.). We refer the reader to the original papers for more details on \texttt{CRASH}.

The spectral shape of the sources is discretized into 82 frequency bins, where 72 are in the UV regime below $h_{\rm P} \nu = 200$ eV, and the rest are in the soft X-ray regime (0.2--2 keV).
    The bins are spaced more densely around the ionization thresholds of hydrogen (13.6~eV) and helium (24.6~eV and 54.4~eV).

The radiation emitted by each source $i$ at a redshift $z$ is dicretized into $N_\gamma$ photon packets.
Each packet, $N_{ {\rm ph}, i}(\nu,z)$ (in units phots Hz$^{-1}$)
holds the total number of photons emitted by source $i$ during the timestep $\Delta t_{\rm em}(z)$ (in s) in the different frequency bins:
\begin{equation}
    N_{ {\rm ph}, i}(\nu,z) = \frac{S_i(\nu,z)}{L_i(z)}f_{\rm esc}(\nu)\varepsilon_i(z) \Delta t_{\rm em}(z),
    \label{eq:photons_bin}
\end{equation}
where $f_{\rm esc}(\nu)$ is the frequency-dependent escape fraction, $\varepsilon_i(z)$ is the rate of ionizing photons the source emits (in phots s$^{-1}$), $S_i(\nu,z)/L_i(z)=\hat{S}_i(\nu,z)$ is the normalized spectral shape (in Hz$^{-1}$), where $S_i(\nu,z)$ defines the Spectral Energy Distribution (SED; in units of ergs Hz$^{-1}$ s$^{-1}$) and $L_i(z)$ the  luminosity (in units of ergs s$^{-1}$).
We adopt an escape fraction\footnote{\texttt{CRASH} allows us to specify the escape fraction in different frequency bands individually for each source.} of UV photons $f_{\rm esc}= 15\%$ for all sources except BHs, for which photons at all frequencies are assumed to escape. Although the adoption of a single value for the escape fraction is an over simplification as in reality it depends e.g. on the mass of dust and gas and redshift of the host galaxy, its density distribution, the mass and location of the stellar sources (see e.g. \citealt{Ciardi2005} and its updated version on arXiv), $f_{\rm esc}$ is to a large degree an unconstrained parameter at low (see e.g. \citealt{Matthee2017} and \citealt{Vanzella2016}) as well as high redshift, where physical conditions could have promoted large values of $f_{\rm esc}$ (see e.g.~\citealt{Kitayama2004,Yoshida2007,Safarzadeh2016} or the recent $z=4$ observations by \citealt{Vanzella2017}).
In a companion paper, focused on the Epoch of Reionization,
we will address the effect of different choices of the escape fraction on the global emissivity and reionization process. It should be noted that cells hosting sources are treated as any other cell, i.e. their ionization and temperature evolution is dictated by the properties of the crossing rays.

The rate of ionizing photons, or emissivity, can be written as:
\begin{equation}
    \varepsilon_i(z) = \int\limits_{\rm 13.6 eV}^{\rm 2 keV} \frac{S_i(\nu,z)}{h_{\rm P}^2 \nu} \diff(h_{\rm P}\nu).
    \label{eq:emissivity_definition}
\end{equation}
For simplicity, from now on, we will omit the explicit dependences on $\nu$ and $z$.

Each of these packets is followed as it traverses the simulation volume, the photons in different frequency bins ionize and heat the intervening hydrogen and helium, and finally it either goes extinct or reaches the cosmic volume boundary\footnote{In our reference simulations we do no use periodic boundary conditions, although this option is available.}. Due to the large simulated volume, photon red-shift is also accounted for.
Our reference cases use $N_\gamma=10^5$ (see Appendix~\ref{app:convergence} for a convergence analysis).

In the following section we will describe how the spectrum, luminosity and emissivity of the sources are modeled.

\subsection{Sources of ionizing radiation}
\label{ssec:source_modelling}
\begin{figure}
    \centering
    \includegraphics[width=\columnwidth]{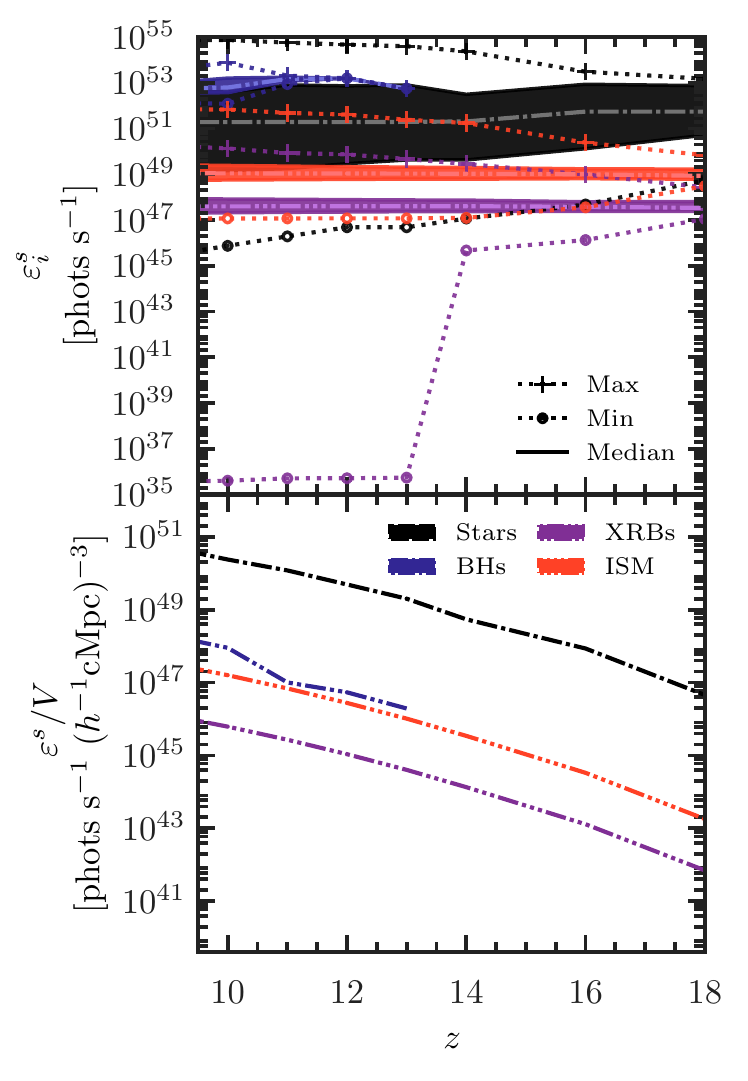}
    \caption{\textit{Upper panel}: Redshift evolution of the emissivity $\varepsilon_i^{s}$ of individual sources (upper panel), grouped per source type $s$.
    Stars are plotted in black (single dot), BHs in blue (two dots), XRBs in purple (three  dot) and the ISM in red  (four dots).
    The solid, central lines are the median values, enclosed in shaded regions that show the 25th to 75th percentile of the values.
    The span between the minimum and maximum emissivities are given from the dashed lines, with ``+'' and ``o'' denoting upper and lower limits, respectively.
    Note that these values are intrinsic and have not been scaled by an escape fraction.
    \textit{Lower panel}: Redshift evolution of the volume averaged emissivity $\varepsilon^{s}/V$ per source type $s$. Here, the escape fractions are accounted for.
}
    \label{fig:emissivity_span}
\end{figure}

In our simulations we consider four types of ionizing sources:
\begin{enumerate}[(i)]
    \item regular stars, hereafter abbreviated only as \textit{stars},
    \item neutron star/black hole X-ray binaries, hereafter \textit{XRBs},
    \item thermal bremsstrahlung from supernova-heated ISM, hereafter \textit{ISM}.
    \item accreting nuclear black holes, hereafter \textit{BHs}.
\end{enumerate}
For each source $i$ of any of the types $s$ we model its SED, $S_i^s$, and luminosity, $L_i^s$, as detailed in the following subsections.
When several sources of different types are present in the same cell,
we sum up their contributions (except for BHs, which are treated separately) to obtain $S_i=\sum_s S^s_i$ and $L_i=\sum_s L^s_i$. Using Eqs.~\ref{eq:emissivity_definition} and~\ref{eq:photons_bin}, we can then evaluate the photon content of each packet emitted.

In Fig.~\ref{fig:emissivity_span} we plot the evolution of the median, the 25th to 75th percentiles and the minimum/maximum values
of the emissivity for the various source types,  as evaluated from Eq.~\ref{eq:emissivity_definition}.
In the lower panel, we plot the comoving volume averaged emissivities. These reflect the abundance of the sources, unlike the individual emissivites plotted in the upper panel, showing that stars provide the bulk of the ionizing photons throughout the redshift range considered here. 
Even though individual BHs have higher emissivities than the majority of the stars in galaxies,
they are much fewer in number, and are therefore not dominating the overall emissivity budget.
The median values of the emissivities for the different source types remain also fairly constant.
This applies to the maximum values too: the brightest stellar-type sources at $z=14$ are less than an order of magnitude brighter at $z=10$.
The brightest black holes, on the other hand, do not have higher emissivities compared to the brightest galaxies, which can emit an order of magnitude more ionizing photons.
The slow evolution in emissivities is related to the physical properties (e.g. stellar masses, star formation rates and metallicities) that goes into determining the SEDs.

We note that the median stellar emissivity decreases with decreasing redshift. 
Although the median stellar mass increases, suggesting higher luminosities,
this is counteracted by an increment of both the median stellar mass weighted metallicity and age of the stars.
Nevertheless, stars are the dominant producer of ionizing photons at all redshifts.
The evolution of the emissivities of the XRBs and the ISM is dictated by the slow evolution of the star formation rates.
The sudden drop in the lower limit of the emissivity of XRBs is due to galaxies that only hosts the fainter LMXBs.
The first black hole arises at $z=13$, and not until $z\leq 11$ do we have more than one present in our volume.

While \texttt{CRASH} can handle a different spectrum for each single source, for the sake of simplicity we adopt at each redshift $z$ an average spectral shape for all sources but BHs. 
More specifically, for each source $i$ we evaluate $S_i=S^{\rm stars}_i+S^{\rm XRB}_i+S^{\rm ISM}_i$. 
We then use the average $\bar{S}=\langle S_i\rangle$ in Eq.~\ref{eq:photons_bin} rather than $S_i$.
Whenever a BH is present in a cell, this is added as a separate source having the same spatial coordinates.
Its spectrum is similarly calculated as $\bar{S}^{\rm BH}=\langle S^{\rm BH}_i\rangle$.

\begin{figure}
    \centering
    \includegraphics[width=\columnwidth]{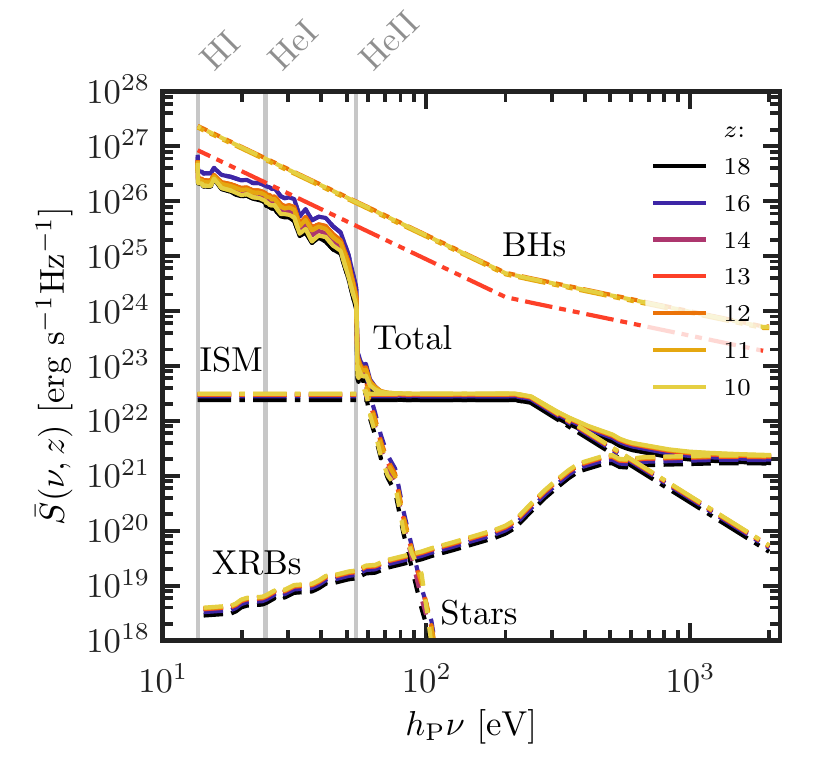}
    \caption{Globally averaged spectral energy distributions (SEDs) $\bar{S}(\nu,z)$ of the different source types: stars (short dashed
    lines), XRBs (long dashed), ISM (short-long dashed), total galactic (solid) and black holes (long/double short dashed).
    The faint, vertical gray lines indicate the ionization thresholds for hydrogen (13.6 eV), neutral helium (24.6 eV) and singly ionized helium (54.4 eV).
    Different colors indicate different redshifts.
}
    \label{fig:galactic_SED}
\end{figure}

In Fig.~\ref{fig:galactic_SED} we plot $\bar{S}$ at different redshifts $z$. 
We see that 
stars dominate at energies $h_{\rm P} \nu \lesssim 60$ eV, while the ISM contribution is relevant above the \HeII ionization threshold, i.e. into hard UV and the soft X-rays.
The XRBs provide the harder X-rays.
The weak redshift evolution is due to a combination of the averaging effect effectively preferring brighter sources, and the mentioned slow evolution in the underlying physical properties determining the spectra.

In the following sections we will describe in more detail how we evaluate the luminosity $L^s_i$ and normalized spectral shape $\hat{S}^s_i$ for the various source types.

\subsubsection{Stars}
\label{ssec:stars}

We model the ionizing radiation from stars by using stellar particles identified in \MBII. 
From the age, metallicity and mass of each stellar particle $p$, we obtain $\hat{S}^{\rm stars}_p$ and $L^{\rm stars}_p$ using the stellar population synthesis code BPASS \citep{Eldridge2012}. 
We adopt the instantaneous starburst prescription of star formation and the evolution model that does not account for interactions in binary systems\footnote{The effect of not including the evolution of binary systems is a reduction in ionizing flux.}.

As a single halo $i$ may comprise several particles $p$, in this case we sum up the contributions from all the particles to obtain $L_i^{\rm stars} = \sum_{p \in i} L_p^{\rm stars}$ and $\hat{S}_i^{\rm stars} = \sum_{p \in i} \hat{S}_p^{\rm stars}$.

\subsubsection{X-ray binaries}
\label{ssec:XRBs}

To account for ionizing radiation coming from X-ray binary systems, we combine the galactic properties provided by \abb{MB}{2} with scaling relations from the \cite{Fragos2013a,Fragos2013} XRB population synthesis model, recently updated  by \cite{Madau2017}. 
We can thus capture both the metallicity evolution of the high-mass XRBs (HMXBs) and the age and stellar mass dependence of the low-mass XRBs (LMXBs), as well as the spectral shape evolution in redshift.  

For each stellar particle $p$, we obtain its spectral shape $\hat{S}^{\rm XRB}_p$ 
from \cite{Fragos2013a,Fragos2013} 
and its luminosity $L^{\rm XRB}_p$ as:
\begin{equation}
    L^{\rm XRB}_{p} = L^{\rm HMXB}_{p} + L^{\rm LMXB}_{p}.
    \label{eq:L_XRBs}
\end{equation}
The contribution from HMXBs can be found using Eq.~(3) in \cite{Fragos2013a} with updated coefficients $\beta_{0-4}$ from \cite{Madau2017}:
\begin{align}
    \log\left(L_{p}^{\rm HMXB}/\text{SFR}_i\right) &= \beta_0 + \beta_1 Z_p + \beta_2 Z_p^2 + \beta_3 Z_p^3 + \beta_4 Z_p^4 \notag \\
    &\times \left({\rm erg\, s^{-1}}\, {\rm M}_\odot^{-1} \,\rm yr\right),\quad \text{for}\, Z \in \left[ 0,0.025 \right]
    \label{eq:fragos_3}
\end{align}
where SFR$_i$ (in $M_\odot$~yr$^{-1}$) is the star formation rate of the halo $i$ hosting particle $p$, and $Z_p$ is the metallicity of the stellar particle.

The contribution to the luminosity from the LMXBs from each stellar particle $p$ is found using Eq.~(4) in \cite{Fragos2013}, also with updated coefficients $\gamma_{(0-4)}$ from \cite{Madau2017},
\begin{align}
    \log\left(L_{p}^{\rm LMXB}/M_p\right) &= \gamma_0 
    + \gamma_1 \log(t_p/{\rm Gyr}) 
    + \gamma_2 \log(t_p/{\rm Gyr})^2 \notag \\
    &\phantom{=} + \gamma_3 \log(t_p/{\rm Gyr})^3 
    + \gamma_4 \log(t_p/{\rm Gyr})^4 \notag \\
    &\times \left({\rm erg\, s^{-1}}\, 10^{10} {\rm M}_\odot\right),\quad \text{for}\, t_p \in [0,13.7] \, {\rm Gyr}
    \label{eq:fragos_4}
\end{align}
where $t_p$ (in Gyr) and $M_p$ (in $10^{10} {\rm M}_\odot$) are the age and mass of the stellar particle, respectively.

The particles luminosities and spectral shapes are added as described in the previous section to obtain the corresponding source characteristics.
Finally note that we adopt  coefficients and spectral shapes that are not attenuated from interstellar absorption. 

\subsubsection{Thermal X-rays from hot ISM}
\label{ssec:ISM}

    We also include ionizing radiation from the diffuse ISM of galaxies.
    The spectral shape $\hat{S}_i^{\rm ISM} (\nu)$ 
    is assumed to be that of thermal bremsstrahlung and constant in redshift \citep{Pacucci2014},
    \begin{equation}
        \hat{S}_i^{\rm ISM} (\nu) = 
        \begin{cases}
            C & \text{for } h_{\rm P} \nu \leq kT_{\rm ISM}, \\
            C \left(h_{\rm P} \nu/kT_{\rm ISM}\right)^{-3}
                & \text{for } h_{\rm P} \nu > kT_{\rm ISM},
        \end{cases}
        \label{eq:SED_ISM}
    \end{equation}
    where $C$ is a normalization constant that ensures the correct units of Hz$^{-1}$, and $kT_{\rm ISM}$ is the thermal energy of the ISM in eV.
    From the spectral analysis of observations of the diffuse gas in galaxies by \cite{Mineo2012a}, we use $kT_{\rm ISM} = 240$ eV, translating into a characteristic temperature of the heated ISM of $\sim 10^6$ K.

    Each halo $i$ has an individual luminosity $L^{\rm ISM}_i$, which we evaluate using
Eq.~(3) in \cite{Mineo2012a}:
    \begin{align}
        L_{i}^{\rm ISM} ({0.3-10\rm\, keV}) / \text{SFR}_i &= \left( 7.3 \pm 1.3 \right) \times 10^{39} \notag \\
        &\times \left( {\rm erg\, s^{-1}} {\rm M}_\odot^{-1} \rm yr \right),
        \label{eq:L_ISM_mineo_3}
    \end{align}
    where 0.3-10 keV indicates the photon energy range this relation was obtained for. 
    Note that $L^{\rm ISM}_i$ is corrected to be free of interstellar attenuation. 
    We also rescale $L^{\rm ISM}_i$ to match our frequency band, which has the lower and upper limits of 13.6 eV and 2 keV, respectively.
  
\subsubsection{Accreting nuclear black holes}
\label{ssec:BHs}

To account for the  ionizing photons originated from accretion disks surrounding nuclear black holes, we identify black hole particles in \abb{MB}{2} and use their accretion rates to determine the production of ionizing photons.
The bolometric luminosity of a black hole $i$ is \citep{Shakura1973}:
\begin{equation}
    L_{i}^{\rm BH} = \eta \dot{M}_{i} c^2,
    \label{eq:QSO-bolometric-lum}
\end{equation}
where $\eta$ is an efficiency parameter, $\dot{M}_{i}$ is the accretion rate and $c$ is the speed of light. 
Consistent with the black hole evolution and feedback in \abb{MB}{2}, we choose $\eta = 0.1$.

As a spectral shape, we adopt the observationally derived mean QSO SED of \cite{Krawczyk2013}, which is based on 108,104 QSOs sampled at $0.064 < z < 5.46$.
When no observational data is available between 13.6 eV and 200 eV, this is derived as  interpolation between the mean SEDs for which they have sufficient observations at both higher and lower energies. 
For energies greater than 200 eV, the spectral shape is modeled as a power law,
\begin{equation}
    \hat{S}^{\rm BH}_i \left( h_{\rm P} \nu > 200 \, \text{eV} \right) \propto \nu^{-1}.
    \label{eq:BH_spectral_shape}
\end{equation}
Finally note that no evolution of the SED with redshift is assumed.

\section{Results}
\label{sec:results}

\begin{table*}
\begin{adjustbox}{center}
  \begin{threeparttable}
    \caption{Thermal and ionization state of the IGM at $z=14$, $12$ and $10$ for different combinations of source types. Note that ionization fractions below $10^{-5}$ are denoted with ``$<$''. 
    }
\label{tab:IGM_state}
     \centering\scriptsize
     \begin{tabular}{l  
             c@{\hskip 5pt} c@{\hskip 5pt} c@{\hskip 5pt} c@{\hskip 25pt} 
             c@{\hskip 7pt} c@{\hskip 7pt} c@{\hskip 25pt} 
 c@{\hskip 7pt} c@{\hskip 7pt} c@{\hskip 7pt} }
        \toprule
        {\bf Source type} & \multicolumn{4}{c}{\bf T$^{a}$ [K]} & \multicolumn{3}{c}{\bf $x_\HII$} & \multicolumn{3}{c}{\bf $x_\HeIII$} \\
                     & Median & Volume & Mass & Neutral &
                       Median & Volume & Mass &
                       Median & Volume & Mass \\
                     &        & avg.   & avg. & avg. &
                        & avg. & avg. &
                        & avg. & avg. \\
        \midrule

\multicolumn{11}{l}{\bf $z=14$} \\
        \midrule
Stars & $10$&$16$&$23$&$16$ & $<$&$7.144\times 10^{-5}$&$1.514\times 10^{-4}$ & $<$&$<$&$<$ \\
Stars, BHs & $10$&$16$&$23$&$16$ & $<$&$7.146\times 10^{-5}$&$1.514\times 10^{-4}$ & $<$&$<$&$<$ \\
Stars, XRBs & $10$&$16$&$23$&$16$ & $<$&$7.165\times 10^{-5}$&$1.516\times 10^{-4}$ & $<$&$<$&$<$ \\
Stars, ISM & $10$&$16$&$23$&$16$ & $<$&$7.201\times 10^{-5}$&$1.522\times 10^{-4}$ & $<$&$<$&$<$ \\
Stars, BHs, XRBs, ISM & $10$&$16$&$23$&$16$ & $<$&$7.217\times 10^{-5}$&$1.523\times 10^{-4}$ & $<$&$<$&$<$ \\
Stars, BHs, XRBs, ISM (X=1, Y=0) & $10$&$17$&$23$&$16$ & $<$&$7.273\times 10^{-5}$&$1.533\times 10^{-4}$ & --&--&-- \\
Stars, BHs, XRBs, ISM (X=0.92, Y=0) & $10$&$17$&$24$&$16$ & $<$&$8.304\times 10^{-5}$&$1.720\times 10^{-4}$ & --&--&-- \\
        \midrule
\multicolumn{11}{l}{\bf $z=12$} \\
        \midrule
Stars & $10$&$73$&$142$&$51$ & $<$&$1.627\times 10^{-3}$&$3.456\times 10^{-3}$ & $<$&$<$&$<$ \\
Stars, BHs & $10$&$73$&$142$&$51$ & $<$&$1.630\times 10^{-3}$&$3.459\times 10^{-3}$ & $<$&$<$&$<$ \\
Stars, XRBs & $10$&$73$&$142$&$51$ & $<$&$1.633\times 10^{-3}$&$3.462\times 10^{-3}$ & $<$&$<$&$<$ \\
Stars, ISM & $11$&$74$&$143$&$52$ & $<$&$1.644\times 10^{-3}$&$3.478\times 10^{-3}$ & $<$&$<$&$<$ \\
Stars, BHs, XRBs, ISM & $11$&$74$&$143$&$52$ & $1.004\times 10^{-5}$&$1.652\times 10^{-3}$&$3.486\times 10^{-3}$ & $<$&$<$&$<$ \\
Stars, BHs, XRBs, ISM (X=1, Y=0) & $11$&$75$&$142$&$54$ & $1.381\times 10^{-5}$&$1.688\times 10^{-3}$&$3.545\times 10^{-3}$ & --&--&-- \\
Stars, BHs, XRBs, ISM (X=0.92, Y=0) & $11$&$79$&$149$&$56$ & $1.524\times 10^{-5}$&$1.923\times 10^{-3}$&$3.960\times 10^{-3}$ & --&--&-- \\
        \midrule
\multicolumn{11}{l}{\bf $z=10$} \\
        \midrule
Stars & $11$&$496$&$886$&$214$ & $<$&$2.009\times 10^{-2}$&$3.581\times 10^{-2}$ & $<$&$<$&$1.073\times 10^{-5}$ \\
Stars, BHs & $11$&$497$&$887$&$214$ & $<$&$2.012\times 10^{-2}$&$3.584\times 10^{-2}$ & $<$&$<$&$1.238\times 10^{-5}$ \\
Stars, XRBs & $12$&$498$&$888$&$215$ & $4.442\times 10^{-5}$&$2.014\times 10^{-2}$&$3.586\times 10^{-2}$ & $<$&$<$&$1.311\times 10^{-5}$ \\
Stars, ISM & $16$&$504$&$897$&$219$ & $1.008\times 10^{-4}$&$2.028\times 10^{-2}$&$3.604\times 10^{-2}$ & $<$&$2.915\times 10^{-5}$&$8.005\times 10^{-5}$ \\
Stars, BHs, XRBs, ISM & $18$&$506$&$900$&$221$ & $1.464\times 10^{-4}$&$2.036\times 10^{-2}$&$3.612\times 10^{-2}$ & $<$&$3.092\times 10^{-5}$&$8.408\times 10^{-5}$ \\
Stars, BHs, XRBs, ISM (X=1, Y=0) & $19$&$510$&$876$&$244$ & $1.927\times 10^{-4}$&$2.073\times 10^{-2}$&$3.664\times 10^{-2}$ & -- & -- & -- \\
Stars, BHs, XRBs, ISM (X=0.92, Y=0) & $19$&$558$&$941$&$252$ & $2.111\times 10^{-4}$&$2.374\times 10^{-2}$&$4.100\times 10^{-2}$ & -- & -- & -- \\
\bottomrule
\end{tabular}
\begin{tablenotes}
    \small
    \item[a] The neutral average is calculated by weighting the values by the neutral fraction $x_\HI$ of the cell. More neutral cells then have larger impact.
This quantity is relevant for studies of the 21 cm signal.
\end{tablenotes}
\end{threeparttable}
\end{adjustbox}
\end{table*}

A central question of this study is: can any of our source types drive an Epoch of Heating separate from an Epoch of Reionization? In other words, \textit{is there heating preceding substantial ionization}?
To examine this, we first discuss the qualitative and quantitative (in terms of average physical quantities) impact of the different source types on the IGM and its ionization and temperature histories (\S\ref{sec:IGM_history}). Second, we focus on  the thermal and ionization state of the IGM at $z=12$  and 10, to discuss the details of these processes by using global maps, difference maps and phase diagrams (\S\ref{sec:IGM_z12}).

For the sake of clarity, throughout the manuscript we will use the following nomenclature to define the various ionization states of the IGM: {\it neutral} when $x_\HII < 10^{-5}$, {\it partially ionized} when $10^{-5} \leq x_\HII < 10^{-1}$, {\it highly ionized} when $10^{-1} \leq x_\HII < 0.9$, and {\it fully ionized} when $x_\HII \geq 0.9$.
 
\subsection{Global history}
\label{sec:IGM_history}

\subsubsection{Average ionization fractions and temperatures}

In this section we present both mass and volume average values of various physical quantities, which are summarized in Table~\ref{tab:IGM_state}.

First, we turn to the \textbf{ionization fractions} of hydrogen and helium.
We limit our discussion to $x_\HII$ and $x_\HeIII$, as the behaviour of $x_\HeII$ is very similar to the one of $x_\HeII$. 
From $z=18$ to $z=10$, they increase logarithmically with redshift, and the main driver is the stars, which dominate the emissivity budget.
The scarcity of black holes make them irrelevant on cosmic scales at $z\geq10$.
Similarly, the other source types, albeit being widespread, do not contribute significantly to the statistics as they are far less luminous than stars.
At $z=18$, the volume as well as mass averaged ionization fractions are well below our convergence limits of $10^{-5}$, independently from the combination of source types. This is true also for the median value (which provides insight into the state of most of the IGM, as also discussed by \citealt{Ross2017}), as the vast majority of the IGM is fully neutral. 
At $z=14$, the first redshift shown in Table~\ref{tab:IGM_state}, the median value is still below $10^{-5}$ for all combinations of source types, while the volume and mass averaged ionization fractions now have increased to $\sim 10^{-5}$ and $10^{-4}$, respectively.
At $z=12$ (i.e. 70 Myr later) the median $x_\HII$ is still below $10^{-5}$, except for the case when all sources are present, for which it is $1.0\times 10^{-5}$.
The volume and mass averaged $x_\HII$ are higher ($\sim 10^{-3}$), with slight variations depending on the combination of source types.
At $z=12$, when all source types are present, we thus observe that the IGM is in the beginning stages of a cosmic, partial ionization phase.

Another 100~Myr later, at $z=10$, the IGM has been under the influence of a factor of five more stellar sources, as well as one additional black hole.
The ionization fractions have increased by an order of magnitude, reaching a volume average $x_\HII \sim 2 \times 10^{-2}$ independent of source types, and a mass average that is about twice as large.
The median hydrogen ionization fraction ranges from below $10^{-5}$ (with stars or BHs), to $4 \times 10^{-5}$ with XRBs, to $> 10^{-4}$ with the ISM.
We can understand the small differences in ionization fractions to come from their sensitivity to highly or fully ionized cells, whose prominence is totally dominated by the contribution from stars.
The median, on the other hand is not as prone to biasing by the highly/fully ionized cells.
 When the ISM is present in addition to the stars, helium is also doubly ionized on cosmic scales, albeit only weakly.
At $z=10$ the IGM is predominantly fully neutral when only stars and black holes are present, while the majority of the gas is partially ionized in the presence of more numerous and diffuse energetic sources.

We now investigate the \textbf{temperature} evolution.
At $z=18$ the volume and mass averaged temperatures are 11 K and 12 K, respectively, irrespective of combination of source types.
The median temperature is also comparable, being 10 K.
The above statistics are identical to those we obtain directly from the hydrodynamical simulation \MBII, i.e. they are unaffected by photo-ionization.
At $z=12$, the temperature statistics differ.
The volume (mass) average is $\sim 70$ (140) K, independent of the combination of source types, compared to a gas temperature of 38 (78) K in \MBII.
Weighing the temperature by the hydrogen neutral fraction we find a value $ \sim 50$ K.
The small differences between combinations of source types indicate that stars are the main driver of the evolution of the averaged temperatures, as they dominate the global emissivity budget.
The median temperatures however, are much lower,  around $10$ K.
Having both XRBs and the ISM present in addition to stars and black holes does not raise  the median temperature by $z=12$.

At $z=10$, we have larger differences in the temperature statistics.
This coincides with the widespread partial (but low) ionization we found.
The volume, mass and neutral fraction averaged temperatures are now in the range  (496---506)~K, (886---900)~K and (214--221)~K, respectively, depending on the increasing combination of source types.
The median temperature is however the least biased indicator of the state of the majority of the IGM.
With only stars (and BHs), it is 11 K, but with all source types present, it is 18 K.
This is an order of magnitude higher than what we obtain from the \MBII, where the median temperature at $z=10$ is 6 K.

To summarize, we find that the IGM at $z=12$ is showing its first signs of transitioning into one of two possible states, depending on source populations.
With the most conservative assumption, i.e.~in the presence of only stars (and black holes), the IGM is mostly fully neutral and cold (10 K), and it remains such down to $z=10$.
On the other hand, if XRBs and the radiation from the ISM in galaxies are also accounted for, the IGM becomes mainly partially ionized by $z=10$.
This very low partial ionization is accompanied by a further temperature increase of $\sim 10$ K in all statistics.
However, they also show that different regions of the IGM will have different temperatures.
This in turn can have observable consequences.
Thus, there is a modest Epoch of Heating, and it largely coincides with large-scale low partial ionization from energetic sources.
We now turn to examine the details of this epoch.
 
\subsubsection{Evolution on lightcones}
\begin{figure*}
    \includegraphics[width=\textwidth]{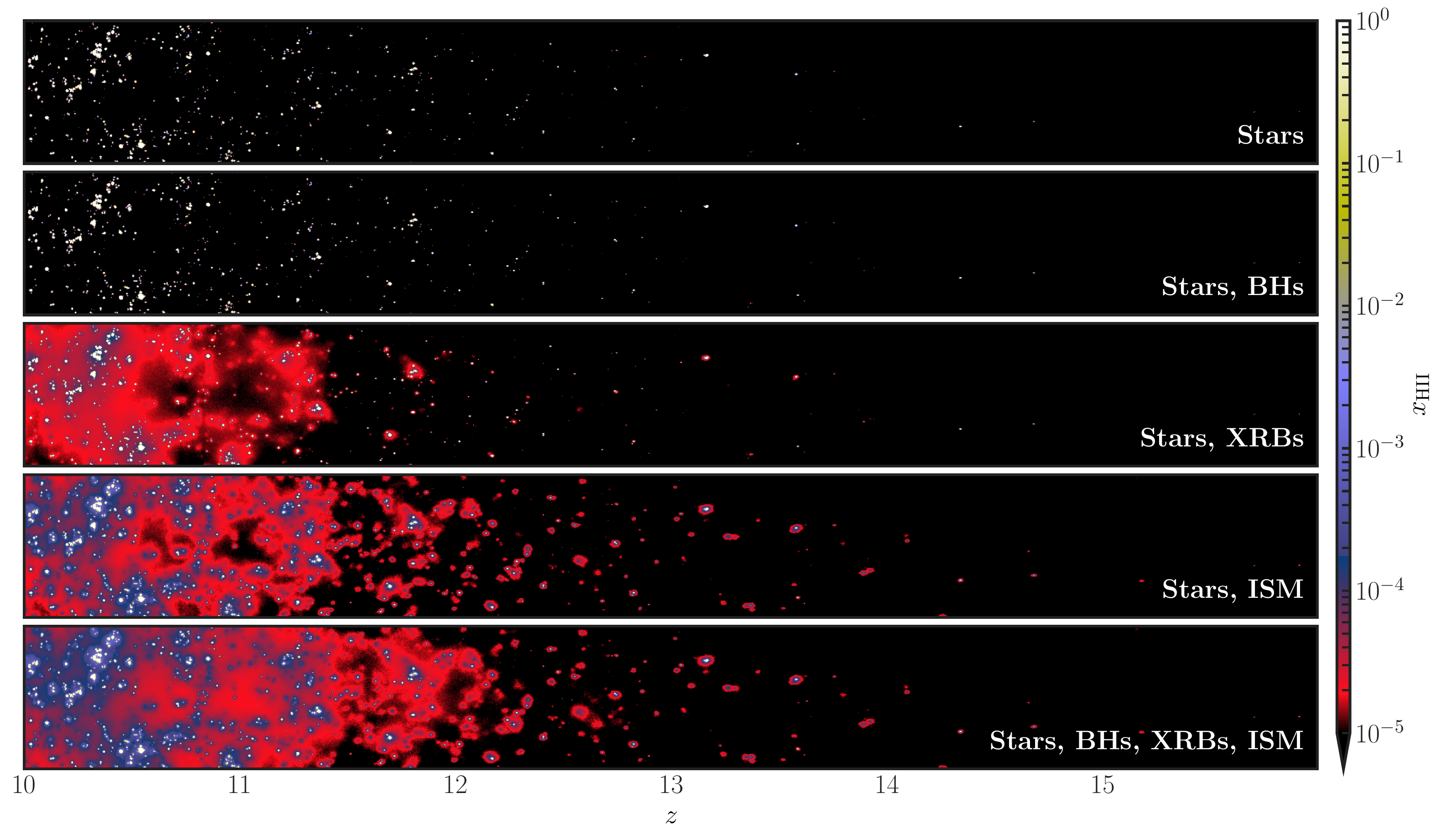}
    \caption{Lightcones showing the evolution of the ionized hydrogen fraction $x_{\rm HII}$ in the simulations with different combinations of source types, as indicated in the labels. The vertical size is 100$h^{-1}$ cMpc and the aspect ratio between the axes is given at $z=10$, hence the evolution in the angular diameter distance is not taken into account. 
    The combined effect of all source types leaves the IGM without fully neutral regions at $z \lesssim 11.5$. 
}
    \label{fig:lightcone_xHII}
\end{figure*}
\begin{figure*}
    \includegraphics[width=\textwidth]{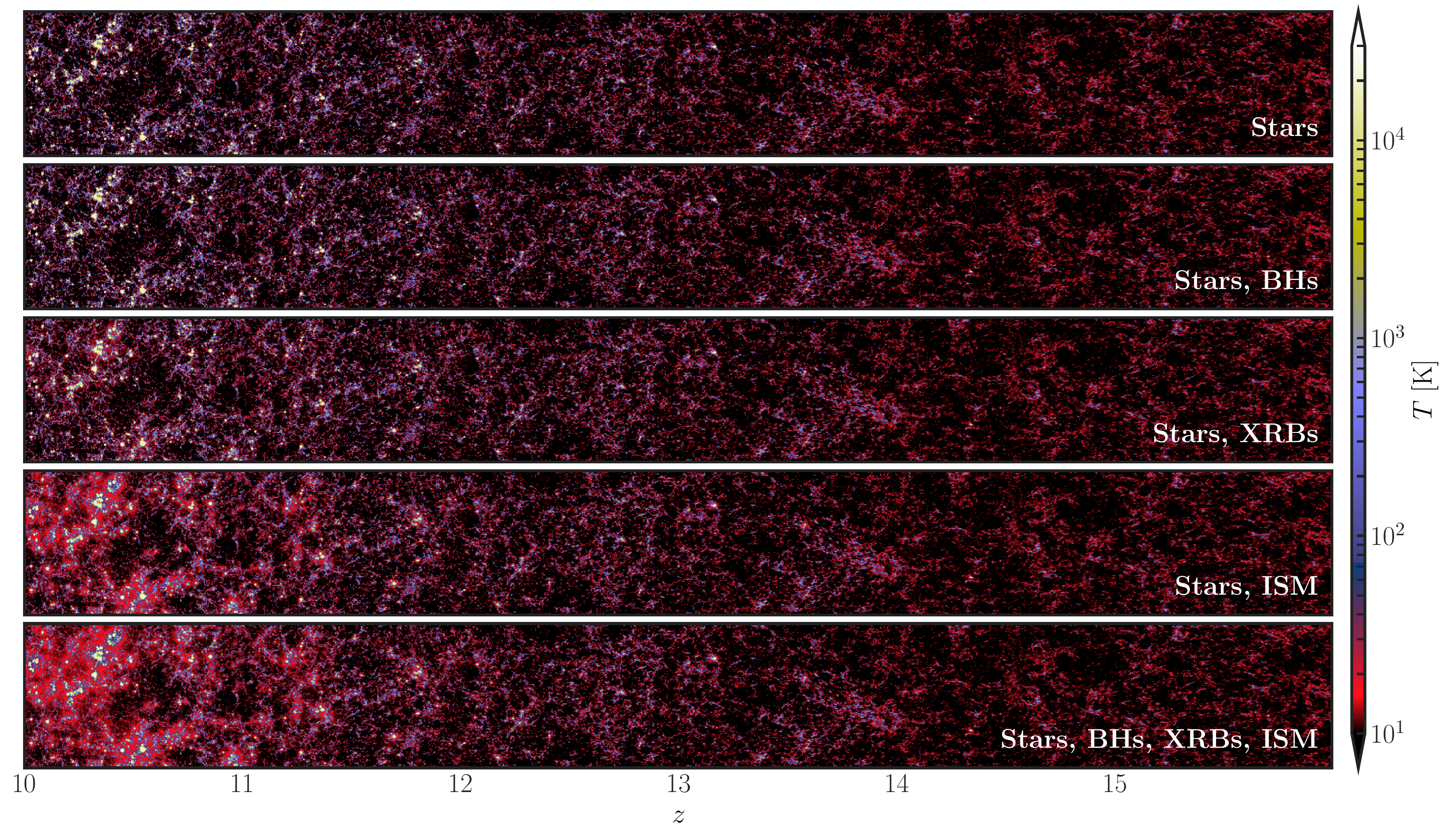}
    \caption{Lightcones showing the evolution of the IGM gas temperature under the presence of different source types as explained in Fig.~\ref{fig:lightcone_xHII}. }
    \label{fig:lightcone_T}
\end{figure*}
\begin{figure*}
    \centering
    \includegraphics[width=0.83\textwidth]{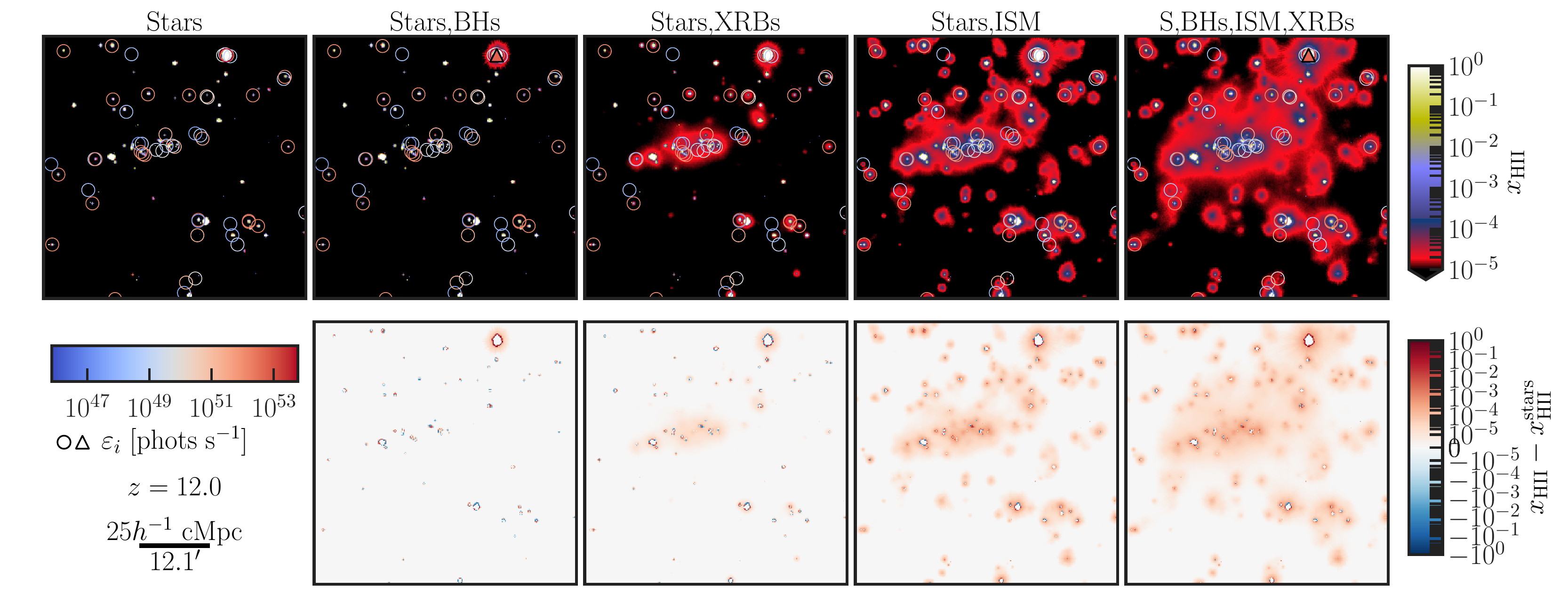}
    \includegraphics[width=0.82\textwidth]{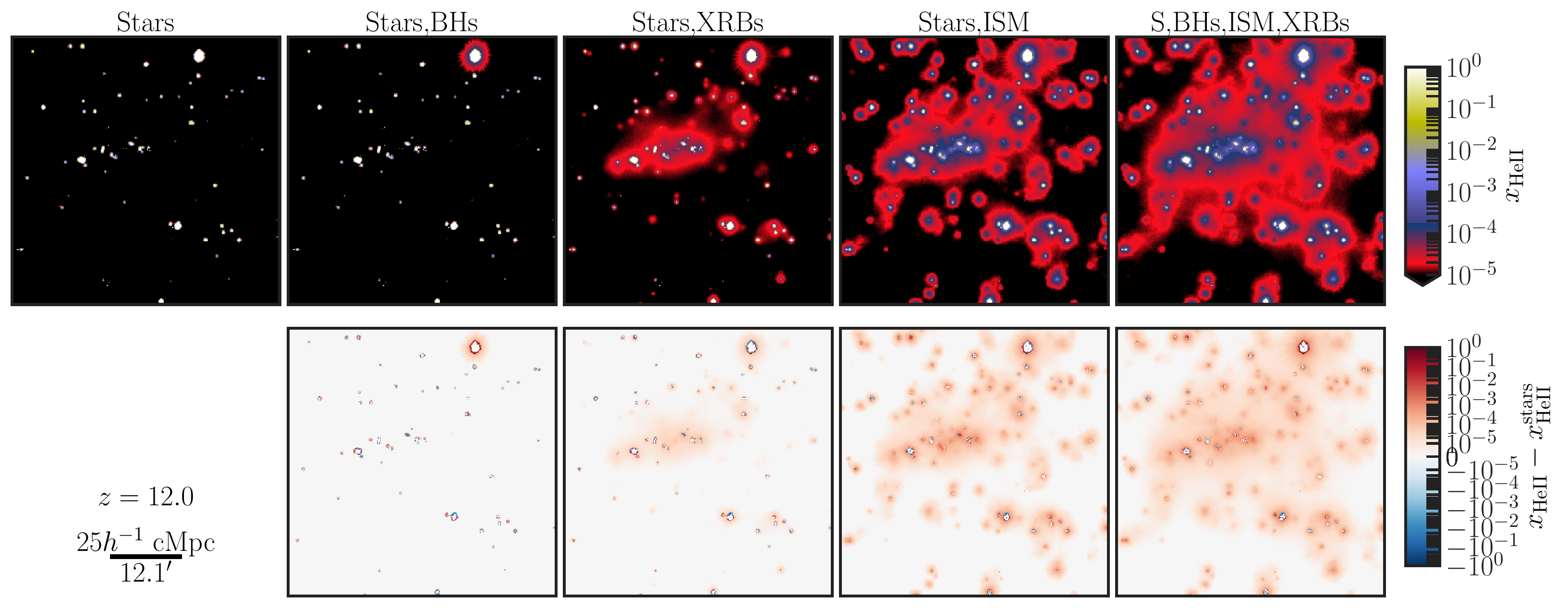}
    \includegraphics[width=0.82\textwidth]{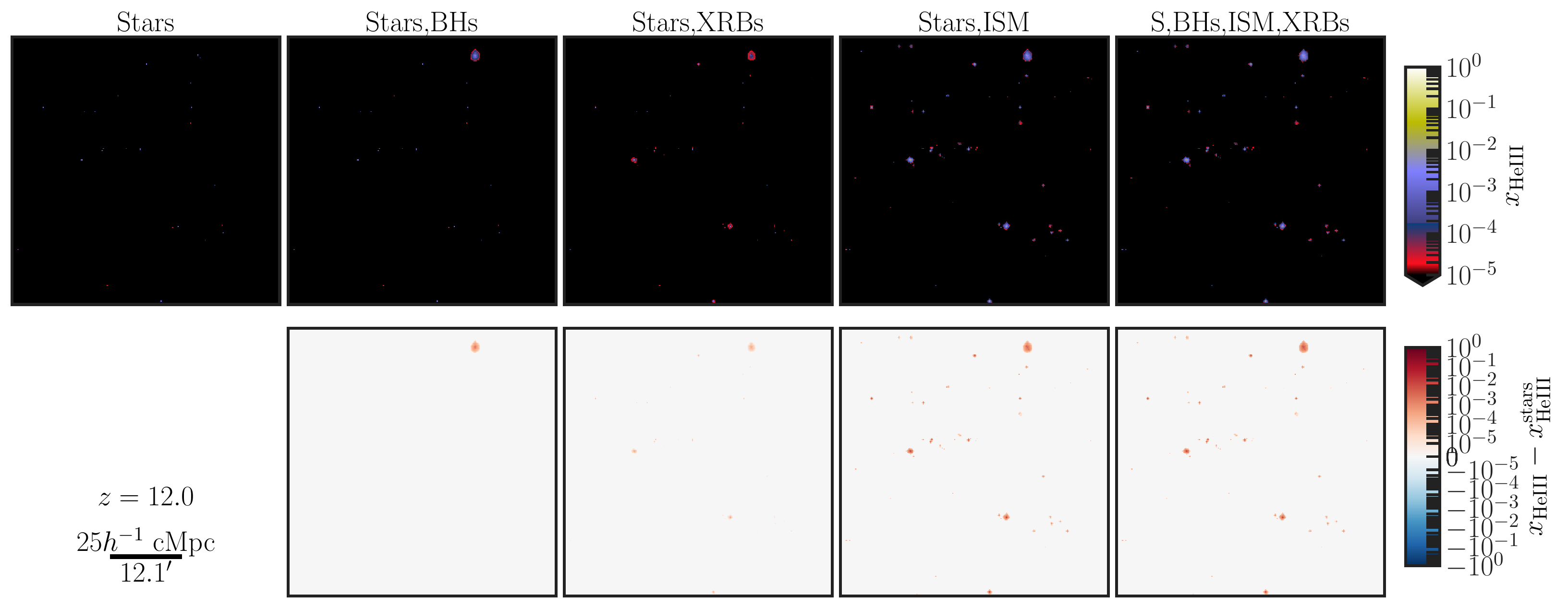}
    \includegraphics[width=0.82\textwidth]{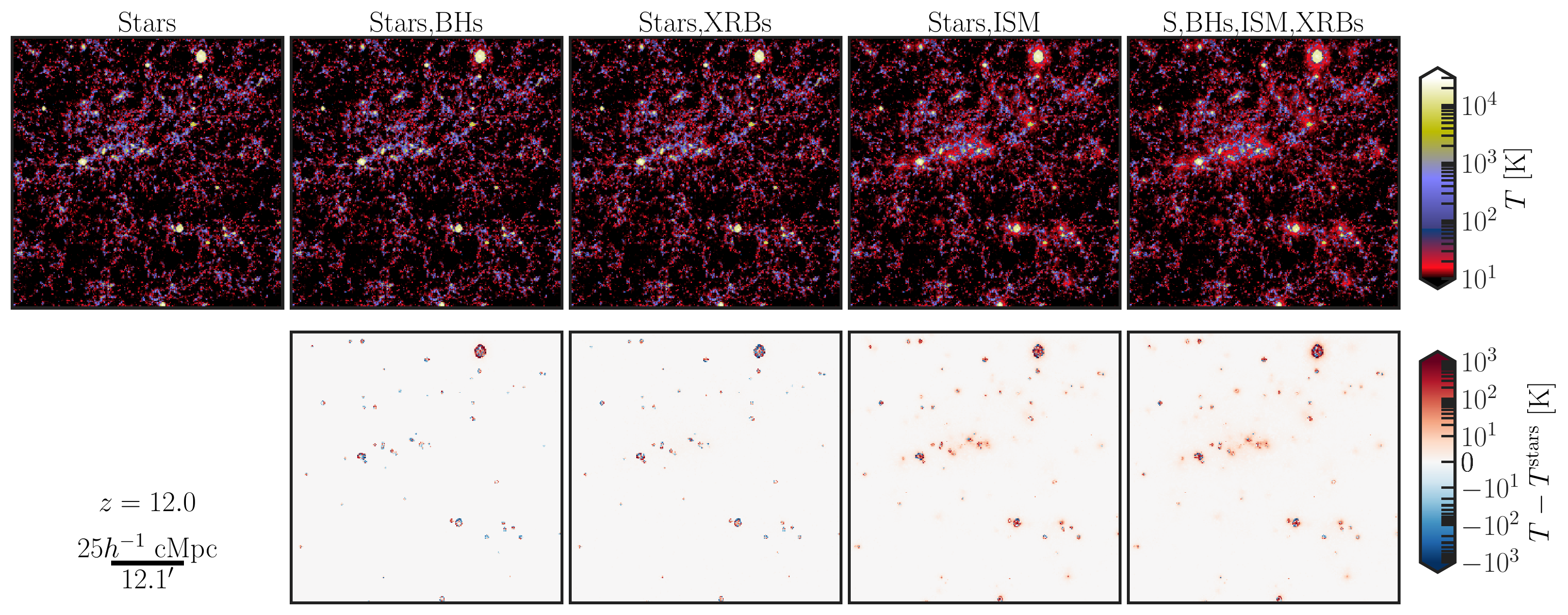}
    \caption{Maps (slices through the volume) in the plane of the only black hole at $z=12$.
             The columns indicate different combinations of source types.
             The upper rows show the quantity in question, and the lower rows show the difference with respect to the first panel, which refer to simulations with stars only.
             In the maps for $x_\HII$, we also show the location of the sources as circles and their emissivities (denoted with the color).
         The only black hole is shown as a triangle in the upper right corner. 
         The maps are $100\h$ cMpc wide.
     }
    \label{fig:difference_maps}
\end{figure*}

To better understand the morphological impact of the source types, in Fig.~\ref{fig:lightcone_xHII} we illustrate how the hydrogen ionization state, $x_{\rm HII}$, evolves in lightcones from $z=16$ to $z=10$. 

We find that stars are determining the existence and shape of the fully ionized regions: clusters of galaxies produce ionization bubbles with sharp ionization fronts.
These bubbles grow mostly separately in our redshift range.
At $z=10$, the IGM is still predominantly neutral, but ionized bubbles are present within distances of tens of cMpc between each other.
With stars we thus have a clear dichotomy with fully ionized bubbles residing in an otherwise fully neutral IGM.

As with the seeding procedure adopted in the \abb{MB}{2} simulation only a handful of black holes is present at $z>10$,
the global history is unaffected by them. 
The first BH appears at $z=13$, while the brightest is at $z=11$ with $M_{\rm AB} = -17.6$.
Its ionizing photon production rate is $\varepsilon^{\rm BH} = 2.05 \times 10^{53}$ phots s$^{-1}$, its mass is $M_{\rm BH} = 1.17 \times 10^6 h^{-1} {\rm M}_\odot$, and it accretes with a rate of $\dot{M}_{\rm BH} = 10^{-2} {\rm M}_\odot$ yr$^{-1}$.
The stars within its host galaxy are even brighter, with $M_{\rm AB} = -18.7$, and produce $\varepsilon^{\rm stars} = 3.22 \times 10^{53}$ phots s$^{-1}$.
In a companion paper focused on the EoR we will investigate the BHs impact on the later stages of cosmic reionization, as well as the effect of a different seeding procedure.

Including more energetic galactic sources (XRBs and the ISM), we note a significant change in the ionization history of the IGM, which becomes partially ionized ($10^{-5} < x_\HII < 10^{-1}$) several cMpc outside the fully ionized regions.
On the other hand, these sources do not significantly contribute to extending the fully ionized regions, which are mainly governed by stars. 
The spectra of the XRBs peak at $\sim 2$ keV \citep{Madau2017}, leaving fewer photons nearer the \HI ionization threshold.
Their long mean free path means they interact after being redshifted significantly, leaving smooth partial signatures of the order $x_\HII \sim 10^{-5}$ tens of cMpc outside their sources.
The spectrum of the ISM is a broken power-law, with a high energy tail
fainter than that of the XRBs, but more photons in the hard UV regime, which are more readily absorbed closer to the source.
With the ISM, we thus see stronger partial ionization gradients, $x_\HII \sim 10^{-3}$--$10^{-5}$, and a more patchy ionization morphology compared to XRBs.

The combined effect of having all source types present in galaxies is shown in the lowermost panel of Fig.~\ref{fig:lightcone_xHII}.
The stars determine the extent of the \textit{fully} ionized regions which, at $z=10$ still make up a very small part of the IGM.
The extent of the \textit{partially} ionized regions goes well beyond those provided by XRBs and the ISM alone.
Combined, they leave a much larger fraction of the IGM in a partially ionized state.

In Fig.~\ref{fig:lightcone_T}, we show the temperature evolution lightcones as well.
The stars effectively determine the temperature of $10^4$ K in the regions we recognize from the previous plot to be fully ionized.
With additional source types, the temperature of these regions does not markedly change, but we do however have heating of the otherwise cold, $T<10$ K, IGM when it is subject to partial and low ionization.
Combined, the XRBs and the ISM provide heating that extends further than either can provide alone, indicating that their combined effect is not simply dominated by the largest of each.
At $z=10$, there is thus a significant difference between the thermal states the IGM can be in, depending on the source types.

\subsection{The IGM at \texorpdfstring{$z=12$}{z=12} and \texorpdfstring{$z=10$}{z=10}}
\label{sec:IGM_z12}
As we have found that the IGM may transition into a predominantly partially ionized, lightly heated state by $z=10$ under the influence of XRBs and the ISM, here we examine its state at $z=12$, when this transition has begun.
This also coincides with the redshift at which we can investigate the effects of the first black hole in our simulation volume, and allows us to quantify its impact, if any.
The reader can refer to Table{~\ref{tab:IGM_state} for some numbers. 

\subsubsection{Morphology}
\label{sec:IGM_morphology}

In Fig.~\ref{fig:difference_maps} we plot maps of (from top to bottom) $x_\HII$, $x_\HeII$, $x_\HeIII$ and temperature at $z=12$ for different combinations of source types. 
These are slices through the volume, chosen to contain  the only black hole in the simulation box, residing in a bright galaxy in the upper right corner.
This black hole accretes with a rate of $10^{-2} {\rm M}_\odot$ yr$^{-1}$ and has a mass of $5.1\h \times 10^{5} {\rm M}_\odot$.
Its absolute magnitude is $M_{\rm AB} = -16.2$ and it has an ionizing emissivity of $\varepsilon^{\rm BH} = 5.6 \times 10^{52}$ phots s$^{-1}$.
It resides in a galaxy where the stars have a higher emissivity, $\varepsilon^{\rm S} = 4.6 \times 10^{54}$ phots s$^{-1}$, while the XRBs contribute with $\varepsilon^{\rm XRBs} = 1.4 \times 10^{50}$ phots s$^{-1}$ and the ISM with $\varepsilon^{\rm ISM} = 4.1 \times 10^{51}$ phots s$^{-1}$.

We also indicate the position and emissivity of the sources with colored circles in the maps of $x_\HII$, while the black hole is represented by a filled triangle.
To highlight the impact of different source types, we also plot absolute differences of the quantities with respect to the simulations with stars only.
Note that we create the difference maps by truncating values below our numerical convergence limit of $x_i = 10^{-5}$ for $i=\HII,\,\HeII,\,\HeIII$ (see Appendix~\ref{app:convergence}).

The stars are responsible for the fully ionized \HII bubbles and their morphology, as we also saw in the lightcones.  \HeII 
follows \HII due to the similar ionization potentials, but as \HeI has an ionization cross section progressively larger than that of \HI at photon energies $h_{\rm P} \nu \geq 24.6$~eV, it is then more readily ionized, and its spatial distribution is generally more extended.
Also note that  sources having in their spectra higher energy photons (see the cases Stars+XRBs and Stars+ISM),  which are redshifted at the simulated scales, provide a contribution  first to  \HeI ionization and then to \HI. These combined effects give us a larger extent of partially ionized $x_\HeII$ than $x_\HII$, while  there is still little overlap between the fully ionized regions, leaving a very patchy \HII and \HeII ionization morphology.
Where sources are more grouped together, they leave imprints on their surroundings similar to what one would expect of separate, but much brighter sources.
The extent of ionization must thus be considered an effect of the ionizing photons arising within a volume, rather than from a single source.
This helps the partially ionized regions to grow considerably in the cases with XRBs or the ISM.

The only BH at $z=12$ accounts for partial ionization in its vicinity, spanning several cMpc, beginning with a smoothing of the otherwise sharp ionization front already created by its host and surrounding galaxies. 
However, a partially ionized tail is not an unique feature of BHs, as it is also seen outside galaxies that have XRBs and the ISM, which both contribute with high energy photons.

As in this epoch, galaxies are more common than BHs, their XRBs and ISM completely dominate the partial ionization at cosmic scales, while the contribution by the ISM component is certainly more ubiquitous than that from the XRBs.
The last panels of both \HII  and \HeII cases finally show the concerted impact of all sources,  resulting in a larger overlap of red areas (\HeII, in particular, is found in the process of merging into a single one) and in a  smoother transition from the fully ionized to the neutral regions. 
\HeII ionization, on the other hand, is much more sensitive to high-energy UV photons ($54.4 \leq h_p \nu \leq 200$~eV), while its cross section at  soft  X-ray energies decreases by 2 orders of magnitude. This is the reason why partially ionized  \HeIII regions with ionizations up to $x_\HeIII \sim 10^{-3}$ are found only within the ionized bubble of the BH or correlating with the brightest galactic sources: the strongest XRBs, and the strongest/most clustered galaxies with a hot ISM contribution.

Maps of the IGM temperature resulting from the various source combinations are finally found in the bottom panels.
The thermal state of the IGM shows signatures both from the hydrodynamic simulation and the heating from the sources.
The ideal gas of the hydrodynamic simulations is heated in overdense regions, whereas shocks occur at scales below the resolution of our grid  ($\sim 400h^{-1}$ ckpc).

As fully ionized \HII and \HeII regions are mainly driven by stars, gas at photo-ionization temperatures of $\sim 10^4$ K strictly correlates with them, as shown in the bottom left panel by white areas. 
Extended blue areas, corresponding to temperatures of $\sim 500$~K, are found along filaments  connecting the stellar sources. These regions, on the other hand, do not have a clear counterpart in the hydrogen ionization pattern because their ionization fraction is below the convergence threshold $x_i < 10^{-5}$.
  
Difference maps show that the addition of other source types does not significantly change the global temperature pattern as setup by the stars. A local increase in $T$ (from 100 up to 1000 K) is found only around the BH and around galaxies having  a hot ISM.
Interestingly, we note that  source clustering still plays a relevant role in changing $T$ but its pattern is less extended compared to the one of the corresponding \HII region. 
Also note that the temperature boost expected in the \HeIII regions from the release of \HeII electrons, as found in \cite{Kakiichi2017}, cannot be appreciated around the first quasars because of the lower emissivity and softer spectral shape.
In fact, their BH had an emissivity $\varepsilon^{\rm BH} = 1.4 \times 10^{56}$ phots s$^{-1}$, and a power law spectrum with spectral index $-1.5$, shifting more ionizing photons closer to the ionization potential of \HeII.

By comparing the relative contribution of different sources we immediately see that XRBs do not heat appreciably the IGM which they partially ionize. We find instead  that the ISM emission, mainly providing softer photons, contributes to a heating of some tens of degrees in the regions that it partially ionizes. These do not extend more than a few cMpc outside the fully ionized regions.
The different impact of XRBs and ISM can be easily explained in terms of their spectral distributions (see Fig.~\ref{fig:galactic_SED}). The energy released by an absorbed X-ray photon is in fact shared between direct photo-ionization, secondary ionization and gas heating. The fraction of the energy that goes into heating increases with the ionization fraction of the gas. 
Heating is more effective for less energetic photons \citep{Dalgarno1999}, and the ISM, which has a softer spectrum, is therefore more effective in heating the gas.

Combining the effect of XRBs and the ISM, we expect a more efficient heating: XRBs should in fact drive pre-ionization in regions that can be influenced by successive contribution of  ISM photons. 
This concerted effect is clearly  visible  as an increase of the extent of the heated  regions created by the presence of all source types. Also check the difference map to have a better feeling of the difference introduced by the concerted, self-consistent, impact of all the source types.

To summarize this section: our models predict an IGM at $z=12$ that is dichotomous, with many, patchy fully ionized and hot regions, which extent and abundance are determined by the stars.
These regions reside in an IGM that is either neutral and cold, or partially ionized when XRBs and the hot ISM of the galaxies irradiate harder UV and X-ray photons.
They can heat the IGM only up to a few tens of degrees, and although this heating is certainly boosted when the XRBs pre-ionize the IGM, it is the ISM that sustains the heating.

\subsubsection{IGM phase statistics}
\label{sec:IGM_phase}
\begin{figure*}
    \centering
    \includegraphics[width=\textwidth]{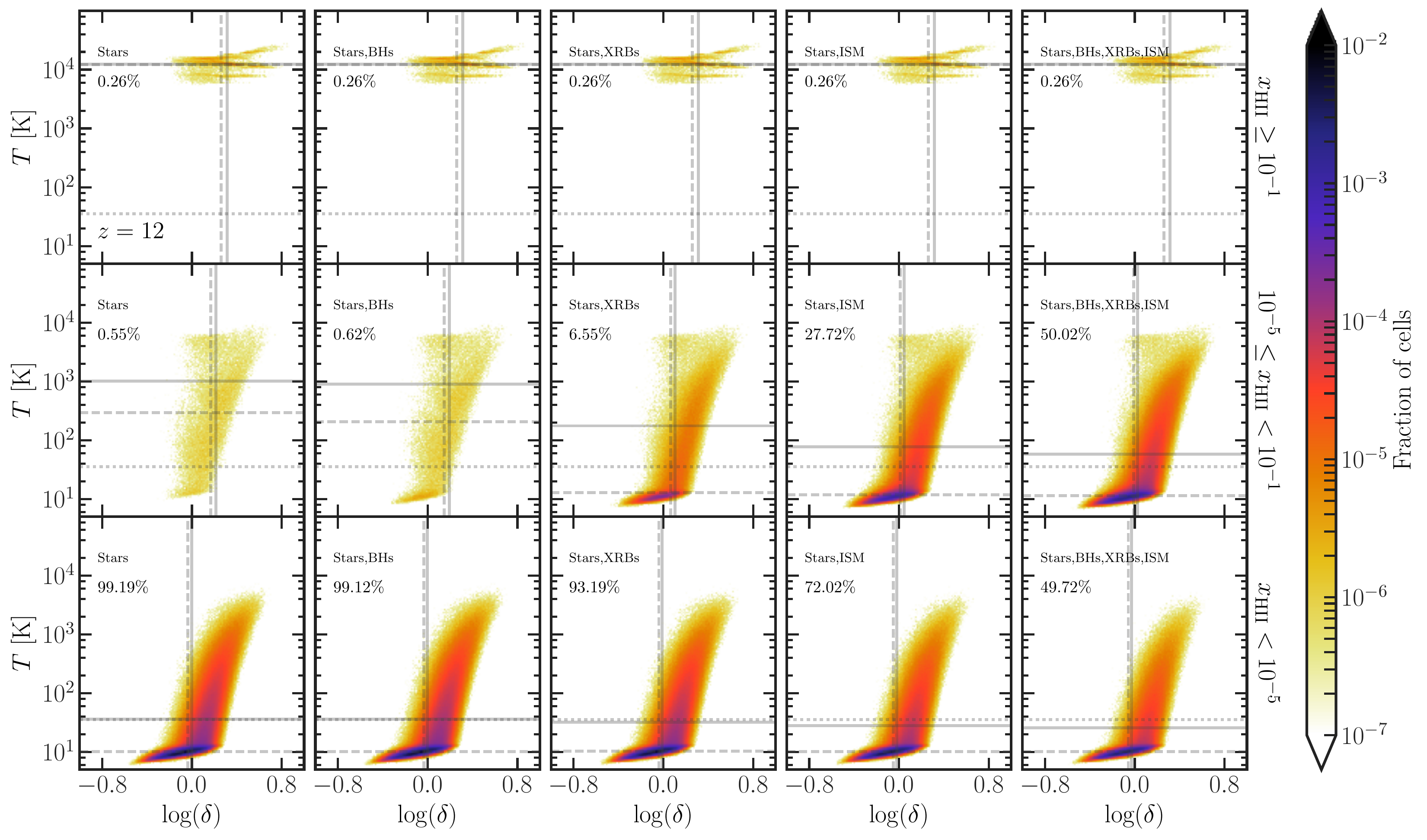}
    \caption{Phase diagrams showing the distribution (in fraction of the total $256^3 =1.68 \times 10^7$ cells) of the IGM at $z=12$ in different thermal ($T$, $y$-axes), overdensity ($\delta \equiv n/\bar{n}$,
    $x$-axis) and ionization ($x_\HII$, rows) states for various combinations of source types, as indicated by the labels. 
    The horizontal solid (dashed) lines indicate the mean (median) temperature in the volume, and the vertical solid (dashed) lines indicate the mean (median) overdensity $\delta$ of the part of the volume with the given ionization state.
    The horizontal dotted lines indicate the CMB temperature.
    The percentages indicate the fraction of the volume that is in the given ionization state.}
    \label{fig:phasemaps}
\end{figure*}

\begin{figure*}
    \centering
    \includegraphics[width=\textwidth]{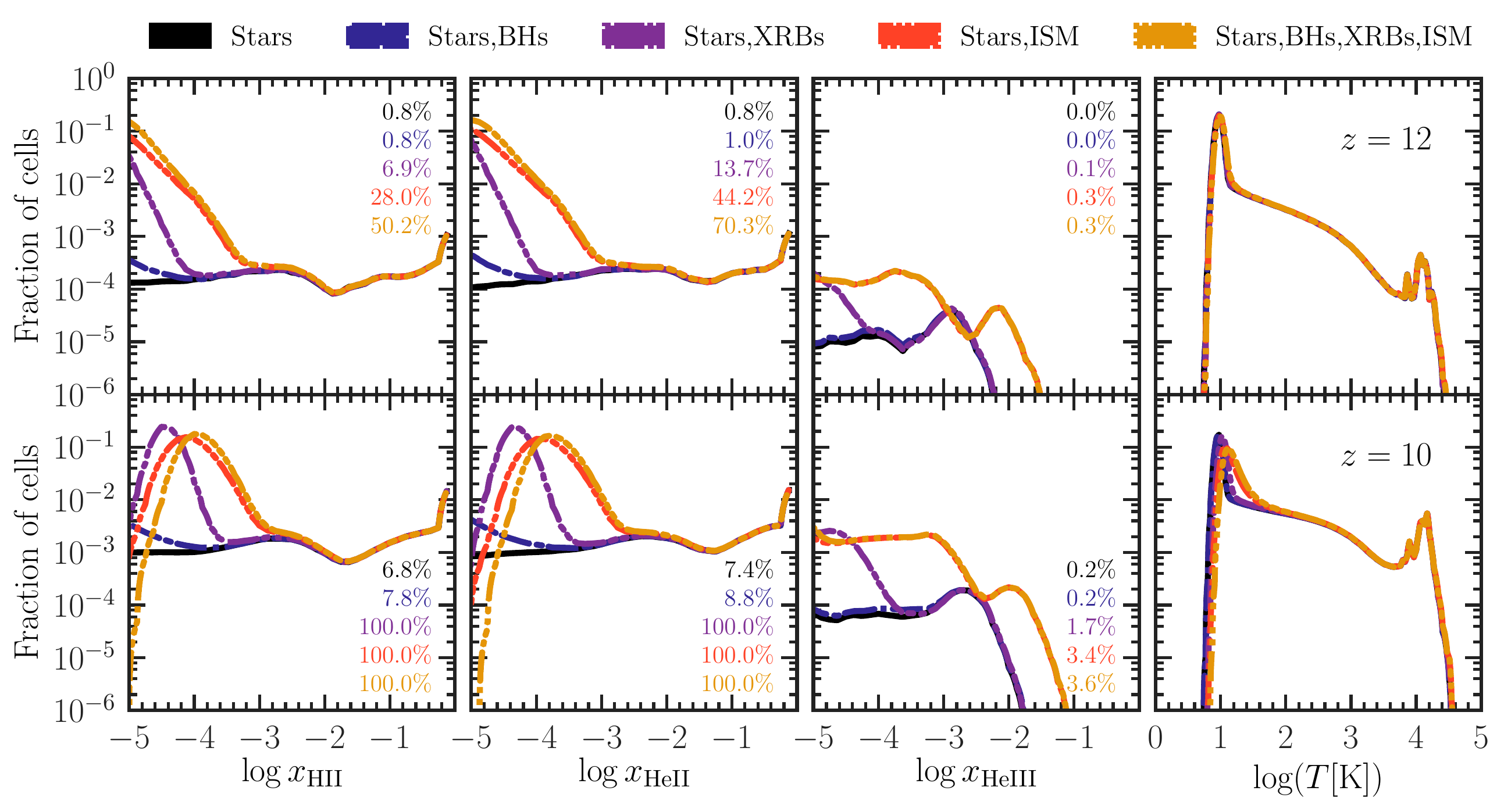}
    \caption{Fraction of cells in different ionization and thermal states for different combinations of source types (\textit{stars}: solid black, no ticks; \textit{stars} and \textit{BHs}: blue, one tick; \textit{stars} and \textit{XRBs}: purple, two ticks; \textit{stars} and \textit{ISM}: red, three ticks; \textit{all}: yellow, four ticks). 
        We show the states at $z=12$ (upper row) and at $z=10$ (lower row).
        The percentages indicate the fraction of the IGM which is shown in the histogram.
        The remaining IGM has an ionization fraction $x_i$ lower than $10^{-5}$ (for $i=\HII,\,\HeII,\,\HeIII$).
    }
    \label{fig:pdf_distributions}
\end{figure*}

In Fig.~\ref{fig:phasemaps} we show the phase state of the IGM at $z=12$ under the influence of different combinations of source types.
Each panel refers to the distributions of the total $1.68 \times 10^7$ cells with a given temperature as a function of overdensity $\delta$\footnote{In this paper the gas overdensity is defined as  $\delta \equiv n/\bar{n}$, where $n$ is the particle number density with units cm$^{-3}$.}. 

The upper row shows the highly and fully ionized IGM to be independent from the type of sources.
The IGM in this state is hot, with mean and median temperatures of $1.2 \times 10^4$ K at a median overdensity $\delta = 2$.
The stars that drive full ionization primarily affect the overdense regions.
As reflected in the values shown in Table~\ref{tab:IGM_state}, a negligible fraction of the gas is in this high ionization state at $z=12$.

The middle row shows the phase state of cells with partially ionized IGM.
We see that a very small fraction is in this state with stars alone, and adding a single BH does not change this.
Including XRBs (ISM) increases the percentage of volume in this partially ionized state to 7\% (28\%), while having all sources present, 50\% of the volume becomes partially ionized.
A comparison with the lower row, which refers to the neutral IGM, clearly show the effect of the additional energetic photons.
The IGM, under the presence of only stars is mostly cold, but with a fraction of cells that is both overdense ($\log \delta > 0$) and hotter than the majority ($T > 10$ K). 
A larger fraction of these cells transitions to a partially ionized state when including the XRBs.
An even larger fraction, both from the cold and underdense as well as the overdense, hotter gas, becomes partially ionized under the influence of the ISM photons.
There is no apparent preference in terms of $\delta$ or $T$ on the IGM that becomes partially ionized. This results from a combination of factors such as the location (overdense regions), emissivity (this is lower for the XRBs than for the ISM) and spectral shape (the XRBs have spectra harder than the ISM) of the sources.
More than 99\% of the IGM is fully neutral with stars only, while this fraction is strongly decreased with XRBs (93\%), the ISM (72\%) or all source types (50\%).

It should be noted that, as already pointed out by other authors \citep[see e.g.~][]{Ross2017}, the partially ionized and warm cells found in the presence of stellar type sources might only arise from a lack of spatial resolution.
More specifically, whenever the cell size is too large to resolve the sharp ionization front expected from stellar type sources, the cell containing the front appears partially ionized and warm, while in reality part of the gas in the cell should be neutral and cold, and part fully ionized and hot. We will discuss this issue in more detail in a companion paper.

In Fig.~\ref{fig:pdf_distributions}, we present histograms of the temperature and ionization fractions at $z=12$ and $z=10$, showing a quantitative evolution of the physical state of the IGM.

As a baseline, we compare the distributions to the case with stars alone, where the curves for $x_\HII$ and $x_\HeII$ are almost flat, with only a surplus of cells in a highly or fully ionized and hot ($T \sim 10^4$ K) state, while the large majority is neutral and cold ($T \sim 10$ K), both at $z=12$ and $z=10$.
The fraction that is highly/fully ionized and hot has increased an order of magnitude by $z=10$.
With the addition of other source types, partial ionization is strongly increased, especially at $x_\HII < 10^{-3}$.
Only at $z=10$ a discernible difference is visible in the temperature distributions, with a shift of the low temperature peak by several tens of degrees when all source types are present.
This is also the redshift where there is largest difference in the fraction of the IGM that has $x_\HII > 10^{-5}$, denoted as percentages in the figures.
At $z=12$, 1\% of the IGM has $x_\HII > 10^{-5}$ when only stars are present, while this fraction has increased to 50\% with all sources.
At $z=10$, this difference is even more significant, as the percentage goes from 6\% to 100\% when including either XRBs, the ISM, or both.
The behaviour of $x_\HeII$ closely follows the one of $x_\HII$.

The distribution and presence of doubly ionized helium changes significantly depending on the source types.
While the presence of XRBs has an impact only at $x_\HeIII < 10^{-4}$, the ISM ionizes more and to higher values of $x_\HeIII$.
At $z=12$, less than 1\% of the IGM has $x_\HeIII > 10^{-5}$.
This fraction, as well as the distribution across all $x_\HeIII$, increases by an order of magnitude for all source types at $z=10$, leaving 4\% of the helium in the IGM in a partially ionized state.

\subsection{Simulations without helium}
\label{ssec:no_helium}
\begin{figure}
    \centering
    \includegraphics[width=\columnwidth]{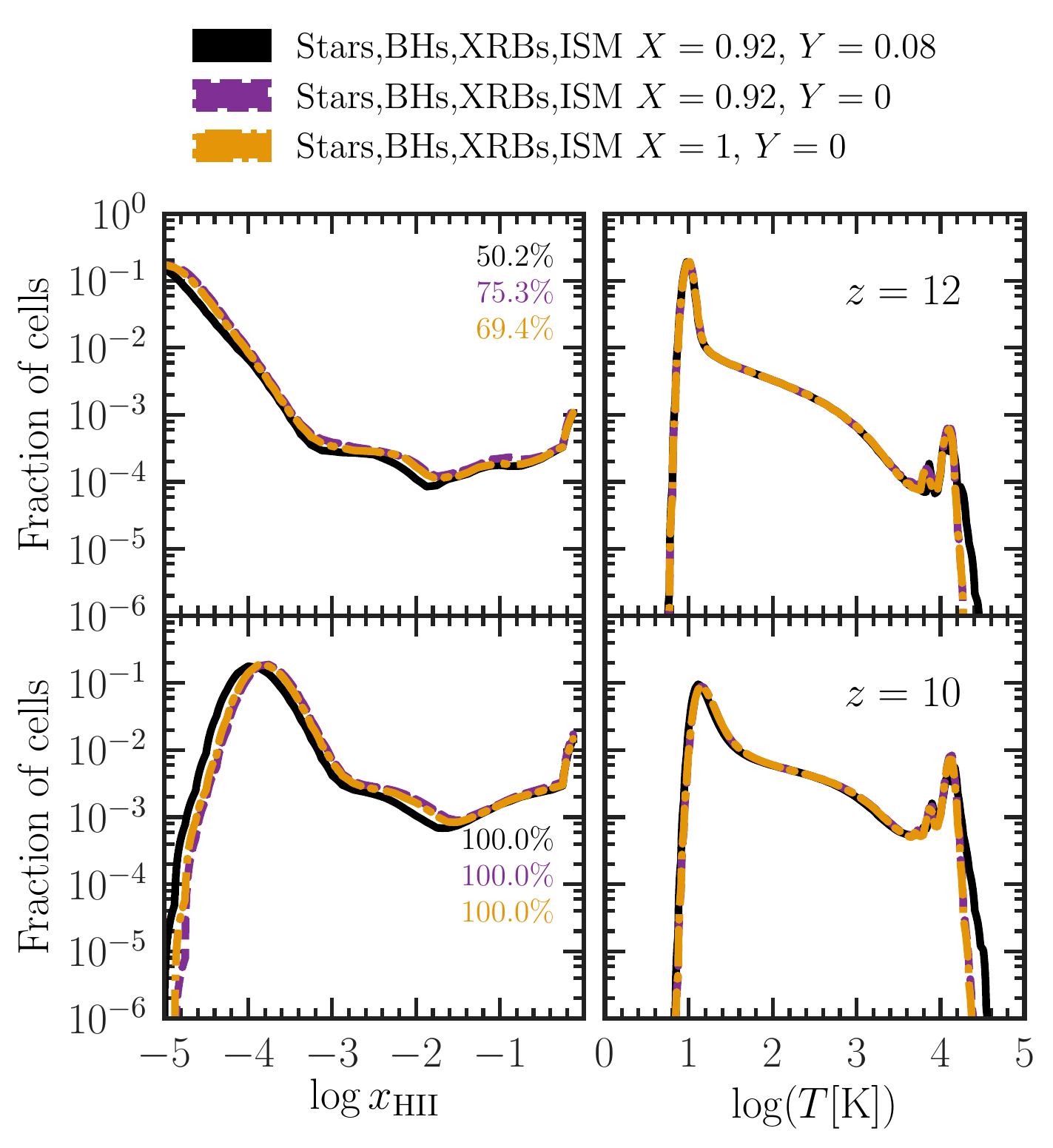}
    \caption{Distributions of volume averaged temperature and ionization fraction of hydrogen for simulations with (solid, black line) or without helium (dashed; purple lines are for $X=0.92$ and $Y=0$, while dash-dotted yellow lines are for $X=1$ and $Y=0$) at $z=12$ and $10$.
    The percentages, with the same colours and reference as the lines, indicate what fraction of the IGM has $x_\HII \geq 10^{-5}$, and hence is partially or strongly ionized.}
    \label{fig:pdf_distributions_noHe}
\end{figure}

The presence of helium in our simulations affects the ionization state of hydrogen as well as the temperature evolution.
The ionization cross section of hydrogen decays roughly as $\nu^{-3}$, and at the ionization threshold of 24.6 eV for \HeI, the cross section of helium is a factor of 6 larger than that of hydrogen.
Similarly, at 54.4 eV, the ionization potential of \HeII, the helium cross section is 16 (neutral) and 13 (singly ionized) times larger than that of hydrogen.
Sources that have spectra that allow for helium to get ionized twice will result in the release of free electrons that can further heat the IGM, as found by \cite{Ciardi2012}.

We have run simulations without helium, i.e. $Y = 0$, including all source types.
We have either replaced the now absent helium with hydrogen, effectively increasing its number fraction from $X = 0.92$ to $X = 1$, or we have simply not solved for helium chemistry, but kept using $X = 0.92$.
Ionization fractions and temperature statistics are presented in Table~\ref{tab:IGM_state}.
When $X=0.92$, we effectively have more photons to ionize hydrogen atoms and heat the IGM, resulting in ionization fractions and temperatures higher at any given redshift compared to the case in which helium is included.
This effect can be mitigated to some extent by increasing the number fraction of hydrogen to $X=1$.

In Fig.~\ref{fig:pdf_distributions_noHe}, we have histograms showing the distribution of $x_\HII$ and $T$ at $z=12$ and $10$.
Dashed and dashed-dot lines indicate simulations without helium.
The increase in ionization fractions mentioned above is accompanied by a uniform shift at $x_\HII < 10^{-1}$ towards higher $x_\HII$ at both $z$, while the distributions of highly and fully ionized cells do not change.
The absence of helium also contributes to another interesting effect: the maximum temperature reached in the IGM is lowered by approximately 0.2 dex,
Without helium, at $z=12$ ($z=10$) the maximum temperature is $T=22,986$ ($36,141$) K, while with helium it is $T=32,313$ ($38,709$) K.
We also have some additional heating of the cold IGM at $z=10$, as seen from the increase in the median temperatures in Table~\ref{tab:IGM_state}.

\section{Discussion}
\label{sec:discussion}

Our simulations show that stars are the principal source of IGM ionization throughout the EoH: in fact they fully dominate the UV part of our spectra (see Fig.~\ref{fig:galactic_SED}) and have a large spatial coverage. 
They are responsible for the volume averaged ionization fractions of both hydrogen and helium, which at $z=10$ are $x_\HII \sim x_\HeII \sim 2.0 \times 10^{-2}$. 
The contribution of stars also drives the shape and extent of local \HII and \HeII bubbles, having sharp ionization fronts due to the short mean free path of their UV photons.
As a rapid phase transformation is not expected until  $x_\HII \sim 0.1$ \citep{Furlanetto2016}, at $z=10$ the IGM is still patchy, with its vast majority being fully neutral, and only a small fraction fully ionized. Only in the presence of more energetic sources, does the morphology and state of the ionized regions change. 

The seeding procedure adopted in MassiveBlack-II, plants seeds of mass $10^5 {\rm M}_\odot$ in halos with $M > 10^{10} {\rm M}_\odot$. This results in a scarcity of BHs at $z>10$, which makes their contribution to IGM ionization and heating negligible. This conclusion, though, is strongly model dependent. We will discuss in more detail the effect of the seeding procedure and the possibility of populating with BHs also smaller halos in a companion paper focused on the EoR.  

It is also important to note that the signature of BHs is degenerate with that of other, more abundant, energetic sources.
The X-ray binaries, accounting for both the sub-dominant low mass binaries (LMXBs) and the SFR-tracing high mass binaries (HMXBs), have spectra that peak at keV scales. They provide a smooth, long-range (several cMpc) partial ionization of hydrogen and helium ($x_\HII = x_\HeII \gtrsim 10^{-5}$). Otherwise, they contribute negligibly, in line with the findings of the semi-analytical works of \cite{Madau2017} and \cite{Sazonov2017}.
Our model of XRBs thus ionizes less than what found by \cite{Ross2017}, who employ a comparable 3D RT post-processing approach, albeit on a dark matter N-body simulation. It should be noted though that their emissivities and source locations were obtained through halo mass scaling relations, which differ from our approach relying on a consistent evolution of baryonic sources (and their properties) from a hydrodynamical simulation.
In particular, by $z=13$ \cite{Ross2017}  find a partial ionization of the cold IGM that is orders of magnitudes higher than what indicated by our computation. They also find  a higher volume averaged $x_\HII \sim 10^{-2}$ independent of whether XRBs were included or not. 
However, while our XRB spectra peak at keV-scales, theirs is a power law  with index $-1.5$.
The luminosity of our XRBs scales with either the SFR and metallicity (HMXBs) or the stellar age and mass (LMXBs) in the halo where they reside, while \cite{Ross2017} assume a constant X-ray production efficiency of each halo.
Our results are more in line with those of \cite{Meiksin2017}, who found the HMXBs to heat the IGM to $T = 22$~K by $z=10$. In their work, though, the increment in temperature from a model with stars only is $\sim 20$~K, compared to our $\sim 1$~K. This could be due to our RT approach capturing only the effect of X-rays up to distances of the order of the box length. On the other hand, \cite{Meiksin2017} do not consider photons with energies below $200$~eV, which are more relevant on smaller scales.
Note also that we are still likely to overestimate the already small contribution from XRBs.
\cite{Das2017}, for example, found the XRB spectra to be attenuated by the ISM, albeit not as strongly as advocated by \cite{Fragos2013a}, but still more than what we have assumed with our constant escape fraction of $f_{\rm esc} = 15 \%$ for photons with energies lower than 200 eV.

Our multifrequency RT highlighted the importance of the ISM contribution as its spectrum provides large abundance of photons in the hard UV and soft X-rays.
We found that this spectral difference is of great importance, a conclusion also shared by \cite{Pacucci2014}, or by \cite{Fialkov2014}, who similarly showed that the hardness of the spectral shape of the XRBs would make them inefficient in heating the IGM.
There is however some uncertainty on the extent of the contribution from the ISM, which 
could be larger, as the supernova-driven heating process of the ISM gas will likely result in emission outside the galactic planes (see e.g.~\citealt{Chevalier1985} or \citealt{Strickland2000}).
In this case, it is less likely that the ionizing UV photons are strongly absorbed by the host galaxy, which is what we assume with our escape fraction.
However, although the production rate of UV photons is uncertain, the ISM is known to supply galaxies ubiquitously with observable amounts of soft X-rays \citep{Mineo2012a}.
The \cite{Meiksin2017} work mentioned above includes also the contribution of the ISM, which raises the IGM temperature at $z=10$ to $T = 6$~K or $T = 34$~K (compared to our median $T=16$~K with stars and the ISM), depending on the wind model adopted. 

One should be careful to attribute long range partial ionization and heating signatures, if observed, solely to the ISM.
These could possibly be degenerate with those from cosmic rays. While \cite{Leite2017} found that they should heat the IGM in the immediate vicinity of star forming halos, this happens because of the cosmic rays confinement. If this confinement is weaker than what estimated by \cite{Leite2017}, their contribution could be very similar to that of the ISM.

Another RT study examining the effect of X-rays on heating and ionization is that of \cite{Baek2010}.
They include X-rays from all of their sources, labeling this contribution as QSOs, but noting that XRBs and supernova remnants also fall into this category.
Their scaling of the X-ray luminosity with the SFR allows for some comparison to our case where all source types are present.
Their mean ionization fraction at $z=10$ is $x_\HII \sim 10^{-2}$ for simulations with helium, but without X-rays.
They note that including QSOs that account for up to 10\% of the total flux does not change this result, similarly to the findings of \cite{Ross2017}, and now also ours.
We have run simulations without helium, finding that the ionization fraction,
contrary to \cite{Baek2010} (cf.~their models S1 and S3) 
increases to $x_\HII = 2.4 \times 10^{-2}$ (keeping $X=0.92$), or $x_\HII = 2.1 \times 10^{-2}$ (with $X=1$).
This is caused by the additional photons that are available for \HI ionization when helium is absent, aligning our results with those of \cite{Ciardi2012}.
Including helium thus provides a slower progression of the cosmic heating and reionization.
As for \HeIII, we find that only faint, diffuse regions having $x_\HeIII \sim 10^{-3}$ start to appear near bright energetic sources.
We do not find the widespread existence of highly/fully ionized \HeIII that \cite{Ross2017} found at higher redshifts.

Consistently with \cite{Madau2017}, we find that XRBs are not able to heat significantly  the IGM due to their hard spectra. Our results show that also the ISM, which has a much softer spectrum, is not able to raise the IGM temperature significantly by $z=10$.
Our median temperatures are in fact $T=11$~K and $T=18$~K for simulations with stars only and all sources, respectively. This is somewhat higher than the value reported in \cite{Baek2010}, where the temperature of the IGM with an ionization fraction $x_\HII < 10^{-2}$ was ranging between 3~K and 7~K without and with QSOs, respectively.
Our volume averaged temperatures are, on the other hand, much lower than those of \cite{Ross2017}, who found $T \sim 600$~K at $z=13$ with stars only and $T \sim 1000$~K with HMXBs, compared to our volume averaged $T = 34$~K at $z=13$ with stars and XRBs.

Predicting the thermal state of the IGM during the EoH is of crucial importance for 21~cm experiments because an IGM with a temperature higher than that of the CMB will be seen in emission, and if lower, it will be seen in absorption.
Interestingly, at $z=10$ the CMB temperature is $T_{\rm CMB} = 30$~K. 
The margins between a 21~cm signal seen in emission or absorption are thus close. 
Note, though, that while our mean temperatures at this $z$ are
much higher than $T_{\rm CMB}$, they do not represent the state of the vast majority of the IGM, which is either fully neutral (with stars only) or partially ionized (with all sources).
We find partially ionized regions that are warmer than the CMB already at $z=12$, as indicated from the neutral and mass averaged temperatures.
We defer to a future work further investigation of the effect that different source types leave on the 21~cm signal.

Observationally, very late (after $z=8.4$) heating of the IGM has been ruled out by measurements of 21~cm line  \citep{Pober2015}.
Less constrained is a period of early and highly efficient heating, which would suppress small scale structure formation by increasing the Jeans mass \citep{Ostriker1996}, disfavouring H$_2$ creation associated with 
X-ray fluxes that do not dominate the ionizing budget
\citep{Oh2001b}, and increase the Thomson-scattering optical depth of the CMB  \citep{Ricotti2004}.

We observe that both the distribution of the sources and their surrounding ionized regions trace the underlying density field.
This can be seen particularly well in the phase maps of Fig.~\ref{fig:phasemaps}, where the fully ionized regions reside where $\log \delta > 0$.
We also see clustering of the sources in the global maps (Fig.~\ref{fig:difference_maps}).
This clustering induces (a) ionized regions that are more extended and (b) heating of the nearby neutral IGM, when comparing the surroundings of sources with similar emissivities but that exist either in a clustered environment or alone.
This distinction (i.e. clustered sources vs single source) is crucial in the interpretation of observations aiming at investigating the physical properties of the high redshift IGM. 

\section{Conclusion}
\label{sec:conclusion}
In this paper we have examined the sources that could end the \textit{Dark Ages}, initiate the {\it Cosmic Dawn}, and drive the \textit{Epoch of Heating} (EoH), when the intergalactic medium (IGM) starts transitioning from a cold and neutral phase, to a hot and ionized one.
The Cosmic Dawn concludes with the \textit{Epoch of Reionization} at $z<10$, which we will explore in a companion paper.
We have investigated how \textit{stars}, \textit{X-ray binaries} (XRBs) and the {\it shock heated interstellar medium} (ISM) of galaxies, as well as \textit{accreting nuclear black holes} (BHs), could heat and ionize the IGM at cosmic scales.
Each source type has spectral characteristics determined by underlying physical processes.
The luminosity of the stars is determined by their masses, ages and metallicities, while the ISM and XRBs are sensitive to the star formation rate in galaxies, as well as their masses and metallicities. 
The black holes, on the other hand, shine with the rate they accrete matter.

We have used the hydrodynamical simulations MassiveBlack-II \citep[\MBII,][]{Khandai2015} to provide us with the physical properties of the sources, their location and abundance, as well as their environment (temperature and baryonic density).
Having established the sources, their spectra and ionizing emissivities, we post-processed \MBII with the radiative transfer code \texttt{CRASH} \citep[e.g.~][]{Ciardi2001,Graziani2013}.

Our main findings can be summarized as follows.
\begin{enumerate}[(i)]
	\item Stars drive the extent, shape, abundance and temperature of the fully ionized \HII and \HeII regions, but do not leave the remaining IGM in any heated or partially ionized state.
    Only 7\% of the overall IGM is in a non neutral state at $z=10$ with stars alone.
    \item With our seeding prescription, nuclear black holes are scarce and do not contribute significantly to heating or ionization at $z>10$.
    \item XRBs contribute to partial, uniform ionization ($x_\HII \sim 10^{-5}$) of the dilute IGM on cMpc scales, but do not significantly heat it by $z=10$. This can be attributed to their hard, keV-peaked spectra.
    \item The ISM of the galaxies contribute to a larger extent of partial ionization than the XRBs do. Their softer spectra provide heating as well as ionization, with $x_\HII \sim 10^{-3}$ and $T\gtrsim 10$ K up to a few cMpc around the fully ionized regions at $z=12$
    and lower partial ionization further out. 
     \item In concert, the ISM and the XRBs induce an an ionization fraction $x_\HII \geq 10^{-5}$ in 50\%  (100\%) of the IGM at $z=12$ ($z=10$). 
    However, at $z=10$ the IGM is still predominantly cold, with median temperatures in the range ($11$--$19$)~K under the influence of stars to all source types.
    \item At $z=10$ the IGM will be seen in both emission and absorption in 21~cm as the neutral averaged temperatures are $T>200$~K for all combinations of source types.
    \item Helium reionization has begun on cosmic scales at $z=10$ in the presence of XRBs and/or the ISM.
\end{enumerate}

If an Epoch of Heating takes place at $z>10$, our findings indicate that it will be a modest one.

\bigskip
\section*{acknowledgments}

We thank Rupert Croft for enlightening comments on the bright black holes; Andrea Ferrara for rewarding discussions; Garrelt Mellema, Philipp Busch, Martin Glatzle, Max Gronke and Aniket Bhagwat for their insights into various parts of the project. MBE thanks Piero Madau for his hospitality.
We thank the anonymous referee for her/his constructive comments.
LG acknowledges the support of the DFG Priority Program 1573 and from the European Research Council under the European Union's Seventh Framework Programme (FP/2007-2013)/ERC Grant Agreement n.~306476.
This work made use of a number of open source software packages, in particular \texttt{numpy} \citep{VanderWalt2011}, \texttt{Cython} \citep{Behnel2011}, \texttt{Matplotlib} \citep{Hunter2007} and GNU \texttt{parallel} \citep{Tange2011a}.


\bibliographystyle{mnras}
\bibliography{references.bib}


\appendix
\section{Convergence}
\label{app:convergence}
\begin{figure*}
    \centering
    \includegraphics[width=0.8\textwidth]{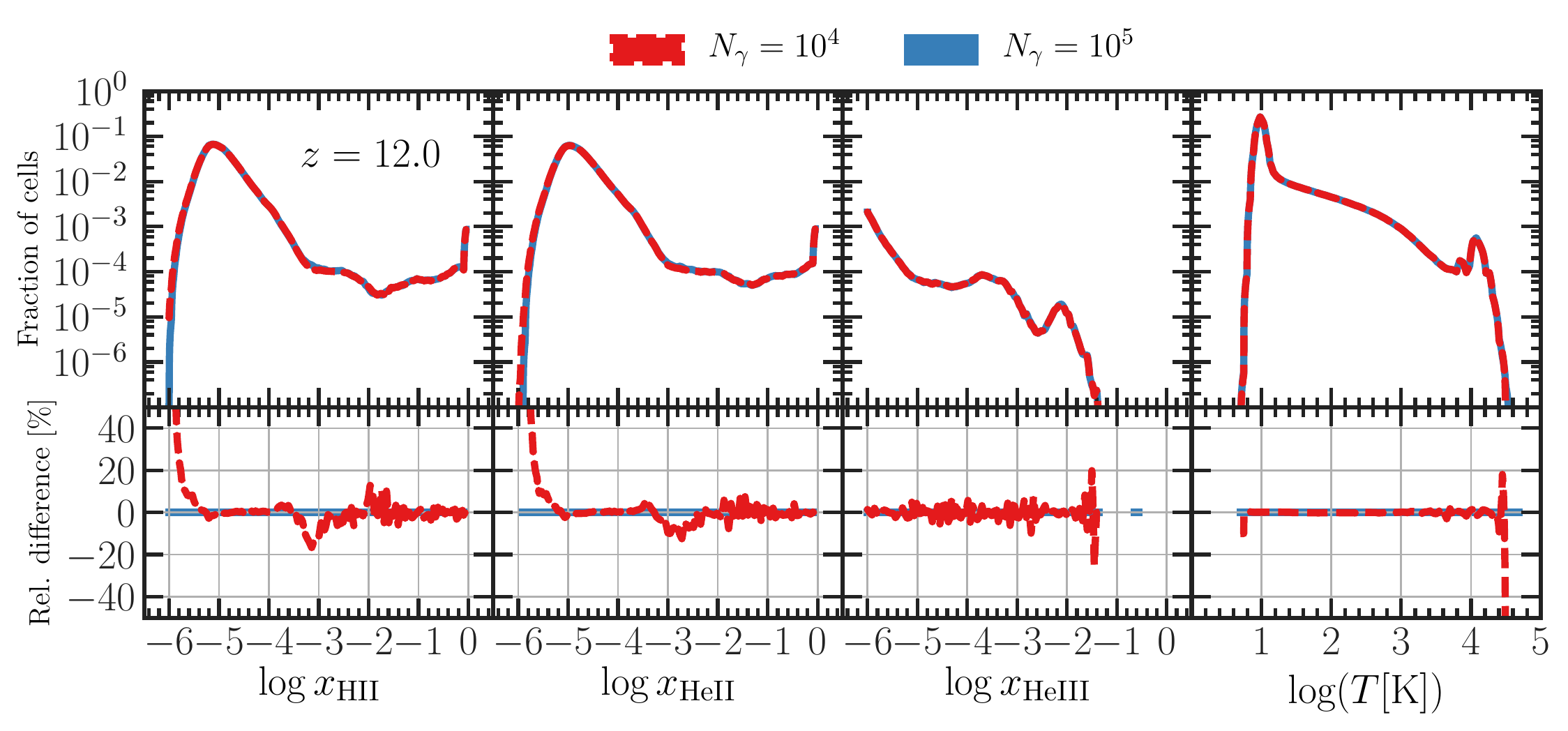}
    \caption{{\it Upper panels}: from left to right, fraction of cells at $z=12$ with a given $x_\HII$, $x_\HeII$, $x_\HeIII$ and $T$ for simulations with $N_\gamma=10^4$ (red dashed lines) and $10^5$ (blue solid lines). 
{\it Lower panels}: relative difference in percent with respect to the case with $N_\gamma=10^5$.}
    \label{fig:convergence}
\end{figure*}

To check the convergence of our results, we run two simulations including all source types which differ only in the number of photon packets emitted per source, i.e. $N_\gamma=10^4$ and $N_\gamma=10^5$. 

In Fig.~\ref{fig:convergence} we show the fraction of cells at $z=12$ with a given $T$, $x_\HII$, $x_\HeII$ and $x_\HeIII$, together with the relative difference between the results of the two simulations, i.e. $D/D_{\rm ref} -1$, where $D={T, x_\HII, x_\HeII, x_\HeIII}$ and $D_{\rm ref}$ represents the values obtained with the highest value of $N_\gamma$.
More than 50\% of the cells in the $N_\gamma = 10^4$ simulation are very well converged, with differences at the percent-level, except for $x_\HII = x_\HeII \sim 10^{-3}$ where the results deviate up to $\sim 20\%$.
For the temperatures, the differences are less than 5\% for the 67\% of the cells that have 10~K $< T \leq 10^4$ K.
 Large differences are observed only in a handful of cells at high temperature or very low ionization fraction $x_\HII = x_\HeII < 10^{-5}$, which we set as the lower limit in this work.
\section{Lightcones}
\label{app:lightcones}
We construct lightcones inspired by the methods applied by \cite{Mellema2006}, \cite{Datta2012b} and \cite{Giri2017}. Our approach though needs to take into account the non periodic boundary conditions of our simulations.
We create a path $\Delta l(t_i) \equiv \Delta l_i$ corresponding to the light travel distance $c \Delta t_i = k_i \Delta x_i$ covered by a ray travelling through $k_i$ cells of the volume, each with sides of physical length 
\begin{equation}
    \Delta x_i = \frac{1}{1+z_i} \frac{100\h}{256} {\rm pMpc},
\label{eq:lightcone}
\end{equation}
from one \texttt{CRASH} output $i$ to the next, $i+1$, having a difference in cosmological age $\Delta t_i$.
We hence assume the time steps to be sufficiently small to neglect the evolution between them when calculating the path $\Delta l_i$.
We linearly interpolate between each output, i.e.~the $k_i$ cells that are covered between two \texttt{CRASH} snapshots have contributions from both snapshots.
The position along the $k_i$ cells that make up the path between two snapshots can be written as $l_p$, where $p = 0,\dots,k_i - 1$.
For each point $l_p$, which corresponds to a position in both physical $(x_p, y_p)$ and redshift $(z_p)$ space, we also have a third spatial dimension, $\mathbf{c}_p(l_p)$, which is plotted as the $y$-axis in the lightcones.
It has contributions from the two snapshots at times $t_i$ and $t_{i+1}$, weighed by the position $(k_i - 1 -p)/(k_i -1)$ and $p/(k_i - 1)$, respectively, along the path between them.
For the ionization fraction we have
\begin{equation}
    \mathbf{c}_p(l_p) = \frac{1}{k_i - 1}\left[\left(k_i - 1 -p\right) \mathbf{x}_\HII^{t_i}(l_p) 
                                 + p \mathbf{x}_\HII^{t_{i+1}}(l_p)
                            \right],
    \label{eq:lightcone_row_p}
\end{equation}
where we write the ionization fractions $\mathbf{x_\HII}^{t_i}(l_p)$ from the two contributing snapshots as vectors to indicate that we only use the cells along the third spatial dimension at this point in $(z_p, x_p, y_p)$-space.

There are various possible approaches for deciding the paths $\Delta l_i$.
Instead of choosing random paths, which could overlap, we instead try fixed patterns, and attempt at covering as much of the simulations volumes as possible in the lightcones.
For this purpose, we choose a continuous whisk broom scanning approach \citep[see e.g.~][]{Schaepman2009} with decreasing horizontal scan width and a several pixels wide (slightly decreasing) vertical scan offset between each horizontal scan. 
Both are reset to maximum widths once we reach the vertical boundary. 
This way we prevent, as far as possible, the same objects at the same cosmological ages to appear in the lightcones.
We only use the interior 236 cells (out of a maximum of 256) along each scanning axis to prevent boundary effects on the lightcones.
The parameters of this approach (the number of pixels to decrease the scan width and offset, vertical scan offset) must however be tuned to obtain a satisfying result.
The approach is thus optimal for a visualization, rather than for an objective quantification, of the temporal progression.

\label{lastpage}

\end{document}